\documentclass[reprint,amsmath,amssymb,aip,jcp]{revtex4-2}

\usepackage{graphicx}
\usepackage{float}
\usepackage{subfigure}
\usepackage{dcolumn}
\usepackage{bm}
\usepackage[colorlinks=true,citecolor=blue, urlcolor = blue, linkcolor= blue]{hyperref}
\usepackage{epstopdf}
\usepackage{color,soul}
\usepackage[toc,page]{appendix}
\usepackage{xcolor}

\begin{document}

\title{Atomistic mechanisms of dynamics in a two-dimensional dodecagonal quasicrystal}

\author{Kun Zhao}
\address{State Key Laboratory of Nonlinear Mechanics, Institute of Mechanics, Chinese Academy of Sciences, Beijing 100190, China}
\address{School of Engineering Science, University of Chinese Academy of Sciences, Beijing 100049, China}

\author{Matteo Baggioli}
\email{b.matteo@sjtu.edu.cn}
\address{School of Physics and Astronomy, Shanghai Jiao Tong University, Shanghai 200240, China}
\address{Wilczek Quantum Center, School of Physics and Astronomy, Shanghai Jiao Tong University, Shanghai 200240, China}
\address{Shanghai Research Center for Quantum Sciences, Shanghai 201315,China}

\author{Wen-Sheng Xu}
\email{wsxu@ciac.ac.cn}
\address{State Key Laboratory of Polymer Physics and Chemistry, Changchun Institute of Applied Chemistry, Chinese Academy of Science, Changchun 130022, China}

\author{Jack F. Douglas}
\email{jack.douglas@nist.gov}
\address{Materials Science and Engineering Division, National Institute of Standards and Technology, Gaithersburg, Maryland 20899, United States}

\author{Yun-Jiang Wang}
\email[Corresponding author. Electronic mail: ]{yjwang@imech.ac.cn}
\address{State Key Laboratory of Nonlinear Mechanics, Institute of Mechanics, Chinese Academy of Sciences, Beijing 100190, China}
\address{School of Engineering Science, University of Chinese Academy of Sciences, Beijing 100049, China}

\date{\today}

\begin{abstract}
Quasicrystals have been observed in a variety of materials ranging from metal alloys to block copolymers. However, their structural and dynamical properties cannot be readily described in terms of conventional solid-state models of liquids and solids. We may expect the dynamics of this specific class of quasicrystalline materials to be more like glass-forming liquids in the sense of exhibiting large fluctuations in the local mobility ("dynamic heterogeneity") and non-Arrhenius temperature dependence of relaxation and diffusion. In this work, we investigate a model dodecagonal quasicrystal material in two dimensions (2D) using molecular dynamics (MD) simulations, with a focus on heterogeneous dynamics and non-Arrhenius relaxation and diffusion. As observed in glass-forming liquids and heated crystals, we observe a two-stage relaxation dynamics in the self-intermediate scattering function $F_s(k,t)$ of our quasicrystal material. It involves a fast $\beta$-relaxation and $\alpha$ relaxation process having a highly temperature dependent relaxation time whose activation energy varies in concert with the extent of string-like collective motion, a phenomenon recognized to occur in glass-forming liquids at low temperatures and crystalline materials at elevated temperatures. After examining the dynamics of our dodecagonal quasicrystalline material in great detail, we conclude that the dynamics of these materials more closely resembles observations on metallic glass-forming liquids than crystalline materials.
\end{abstract}

\maketitle

\section{Introduction}
\label{sec:1}

According to the traditional paradigm of solid-state physics, solids are divided into two types: crystals and amorphous (or non-crystalline) materials. Crystals display long-range order and they are defined by an ordered lattice with translational periodicity. They are compatible only with discrete $1$-, $2$-, $3$-, $4$-, and $6$-fold rotational symmetries and they present sharp Bragg peaks in x-ray diffraction. On the contrary, amorphous solids lack all of these properties, and, in particular, they do not exhibit long-range order. In 1982, Shechtman \cite{Shechtman1984} (see also \cite{Shechtman1985}) discovered a crystal exhibiting $5$-fold symmetry -- a type of ordered structure previously thought to be ``impossible". Despite lacking periodicity, an icosahedral phase with $5$-fold symmetry was shown to exist and to exhibit well-defined and intense Bragg peaks as found for crystals, and distinct from amorphous solids where only broad peaks are observed associated with local ordering in the material. The crystal that Shechtman discovered was the first example of an aperiodic crystal, or as latter termed by Levine and Steinhardt \cite{PhysRevLett.53.2477, PhysRevB.34.596} a ``quasicrystal'', short for quasi-periodic crystal \cite{steinhardt2019second, janssen1988aperiodic, divincenzo1999quasicrystals, janot1997quasicrystals}. It was recognized early by Ishimasa et al. \cite{Ishimasa1985} that quasicrystals represent an intermediate ordered state between crystalline and amorphous matter which is not periodic.

Angell \cite{ANGELL2000791} has argued that many forms of matter qualify as being characterized as ``intermediate'' forms of matter, e.g., globular proteins, plastic crystals, incommensurate structures, twisted 2D heterostructures, and glass-formers undergoing fragile-to-strong glass formation, and he observed that such materials often seemed to share common thermodynamic and dynamic characteristics, even if the fundamental origin of these properties is not currently understood theoretically. In the case of quasicrystals, neutron scattering studies have repeatedly indicated that the properties of quasicrystalline materials and incommensurate structures often resemble those of metallic glasses rather than crystalline materials. This was first noticed for the static structure factor of quasicrystalline materials \cite{PhysRevLett.55.2324}, and then recognized also for dynamical properties, connected to anomalies in the vibrational density of states (vDOS) and the low-temperature heat capacity \cite{PhysRevLett.59.102, PhysRevB.63.214301, hafner1998elementary, PhysRevLett.93.245902, PhysRevB.73.193102, PhysRevLett.114.195502, PhysRevB.99.054305, BILJAKOVIC20121741, bb1, Jiang_2023, Janot1993, Windisch1994, Hafner1995, Bohmer2024, Dyre2024}. We point out that while quasicrystals exhibit long-range orientational order as found in crystalline materials also exhibit a large configurational entropy, as in glass-forming liquids above their glass transition because of the many realizations of the quasicrystalline configurations \cite{Fayen2024}. Indeed, the remarkable thermodynamic stability of the quasicrystal materials has recently been argued to be a direct consequence of this large configurational entropy \cite{Fayen2024}. 
In history, the concept of entropic stabilization of quasicrystals, particularly the random tiling hypothesis, has been explored for nearly 30 years~\cite{Elser1985,Henley1988}.

Given the relatively ordered nature of quasicrystalline materials in relation to metallic glass materials, it is then interesting to take a closer look at the dynamics of these materials and to compare with established trends for metallic glasses and heated crystalline materials to determine if this similarity to glass-forming materials also holds in relation to structural relaxation and atomic diffusion.

Since the discovery of quasicrystals there has been avalanche of scientific activity aimed at discovering real materials of this kind and in theoretically understanding the thermodynamics and dynamic properties of quasicrystalline materials which are now known to occur in many forms. Engel \cite{PhysRevB.82.134206} made a pioneering study of the thermodynamics and dynamics of decagonal and dodecagonal quasicrystals, the two most prevalent types of quasicrystals that occur in near two-dimensional (2D) materials. The dynamic behavior of quasicrystals is typically described by two elementary excitations: phonon and phason. Phonon describes the atomic oscillatory motion around equilibrium positions. 
Phason modes correspond to the local rearrangements of atoms. The phason modes and the associated dynamics involving flips have been studied extensively in the literature~\cite{Socolar1986,Baake2012}.
 At the continuum level, phason modes are diffusive in nature. The elementary process of a phason mode is "phason flip" on the atomistic scale. A phason flip entails the overcoming of an energy barrier, resulting in the transformation of a local configuration into another with a comparable topological structure and proximate energy level \cite{PhysRevB.82.134206}. The  decagonal type of quasicrystal was found to exhibit a relative local  particle displacement dynamics occurring through phason flips, which led to a nearly Arrhenius temperature dependence of diffusion in this type of liquid crystal. However, collective particle exchange motion and non-Arrhenius diffusion were found to be more prevalent in dodecagonal quasicrystals, a class of quasicrystalline materials that is more prevalent in soft matter materials such as polymers than metallic materials. \cite{Noya2021, Gemeinhardt2018}. Damasceno \cite{Damasceno2017} have reviewed numerous observations on these model quasicrystalline materials following their initial study.

This study employs a comprehensive approach to investigate the thermodynamic properties and dynamic behavior of 2D dodecagonal quasicrystal, while establishing a systematic comparative framework to identify structural and dynamical parallels with glass-forming liquids and heated crystals. The experimental claims that quasicrystals resemble glass-forming liquids above really do not answer these questions. A careful analysis of the heterogeneous and collective dynamics of the quasicrystal is required, since as we shall see below, heated crystals and glass-forming liquids have many common features in their dynamics that make answering this question more difficult than one might expect.

Molecular dynamics (MD) simulations provide a suitable tool for elucidating basic thermodynamic and dynamic properties of quasicrystals. In this regard, Engel and Trebin \cite{Engel2007-1} introduced the Lennard-Jones-Gauss (LJG) as a model for 2D quasicrystals. The potential has proven to be able to reproduce several entropically stabilized quasicrystal phases, \textit{e.g.}, see Ref. \cite{PhysRevLett.106.095504}. From an atomistic level, MD simulations also provided early evidence for short-time stochastic particle dynamics in the form of phason-flips \cite{PhysRevB.82.134206}. Interestingly, this can happen either with single-particle jumps, but also correlated multi-particle exchange events \cite{PhysRevB.82.134206, He:ug5027}. This string-like collective dynamics is evidently suggestive of a link between quasicrystals and glasses, where this type of particle exchange motion is ubiquitous.

Since there has been extensive previous work on the LJG model for molecular parameters and temperature values leading to dodecagonal quasicrystalline materials, we briefly summarize thermodynamic properties that will serve as reference point in our novel simulations of the dynamics of this model quasicrystalline material. In particular, Sec.~\ref{sec:2} introduces the LJG-type potential, outlines the phase diagram describing the dodecagonal quasicrystal system and provides other details regarding our computer simulations. In Sec.~\ref{sec:3}, we quantify relaxation mechanisms, dynamic heterogeneity, bulk self-diffusion, and the intricate interplay among these basic dynamical properties. We observe temperature-dependent string-like motions within the quasicrystal. Finally, in Sec.~\ref{sec:4}, we conclude with a integrated discussion of our findings. 

\section{Methods}
\label{sec:2}

\subsection{Lennard-Jones-Gauss potential}

\begin{figure*}
\centering
 {\includegraphics[width=0.9\textwidth] {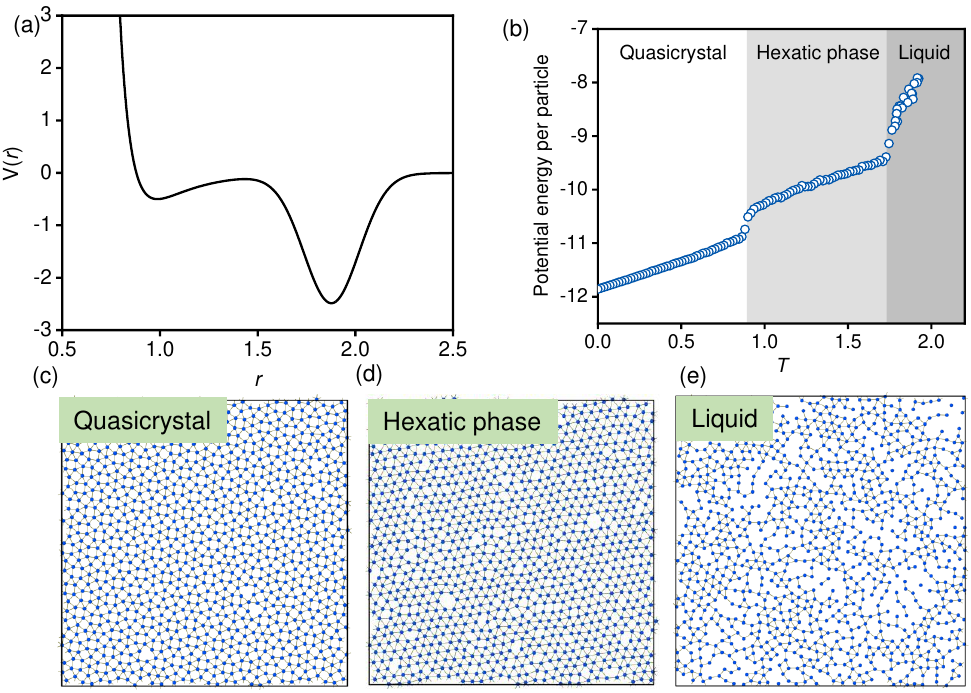}}
  \caption{The LJG-type potential and phase diagram. (a) Double-well nature of the potential with a Gaussian dip. (b) Two phase transitions upon cooling. (c)-(e) Atomic configurations of quasicrystal, hexatic phase and liquid at temperatures $T = 0.3$, $1.2$ and $1.95$, respectively.
  }
  \label{fig:1}
\end{figure*}

We conducted classical MD simulations using the LAMMPS (Large-scale Atomic/Molecular Massively Parallel Simulation) software \cite{Plimpton1995} to investigate the thermodynamic and dynamic properties of a 2D dodecagonal quasicrystal. Our system comprises identical particles interacting via a novel LJG-type potential, which combines the traditional Lennard-Jones potential with a secondary Gaussian well at longer distances. The LJG potential is defined as follows:
\begin{widetext}
\begin{equation}\label{eq:1}
V\left(r\right)=4{\varepsilon_0}\left[{{{\left( {\frac{r}{{{\sigma_0}}}}\right)}^{-12}}-{{\left({\frac{r}{{{\sigma_0}}}} \right)}^{-6}}}\right]+\frac{h}{{{\sigma_{\rm{G}}}\sqrt{2\pi}}}\exp\left[{-\frac{{{{\left({r-{r_{\rm{G}}}}\right)}^2}}}{{2\sigma _{\rm{G}}^2}}}\right].
\end{equation}
\end{widetext}
In this equation, $\varepsilon_0$ is the energy minimum of the LJ well and $\sigma_0$ is defined as the zero-crossing distance for LJ potential. The parameter $h$, in conjunction with the standard deviation $\sigma_{\text{G}}$, dictates the peak height of the Gaussian component, while $r_{\mathrm{G}}$ denotes the peak position of the Gaussian well contribution to $V(r)$. Following previous work \cite{PhysRevB.82.134206, PhysRevLett.106.095504, Gemeinhardt2018}, the chosen parameter values that resulted in the formation of a dodecagonal quasicrystal are $\varepsilon_{\mathrm{0}}=0.5$, $\sigma_{\mathrm{0}}=0.88$, $h=-0.874$, $r_{\mathrm{G}}=1.879$, and $\sigma _{\textrm{G}}^2=0.02$. This choice is made by the prevalence of dodecagonal quasicrystals having this symmetry in real materials, especially soft matter materials such as polymers \cite{Noya2021, Johnston2011, VanDerLinden2012, Lieu2022}. All quantities are expressed in dimensionless Lennard-Jones units. We shall use temperature unit of $\varepsilon_{\mathrm{0}} / k_{\mathrm{B}}$ (here $k_{\mathrm{B}}=1$), energy unit of $\varepsilon_{\mathrm{0}}$, length unit of $\sigma_{\mathrm{0}}$ and time unit of $\tau = \sqrt{m\sigma_{\mathrm{0}}^2/\varepsilon_{\mathrm{0}}}$. A single MD time step is equal to $0.001\tau$. With these parameters, the LJG-type potential exhibits a double-well nature, as illustrated in Fig.~\ref{fig:1}(a). For practical purposes, we applied a cutoff to the LJG-type potential at $r=2.5$.

\subsection{Phase diagram}

A system comprising $N = 1024$ particles arranged in a square lattice is constructed. The system is heated to liquid phase in very fast rate and is fixed at $T = 1.95$ finally. The simulations are conducted with periodic boundary conditions applied in both spatial dimensions. Following an initial period of 50,000 molecular dynamics (MD) steps to ensure equilibration, the system undergoes a controlled linear temperature decrease to reach $T = 0.001$ over a course of $3.9 \times 10^7$ MD steps. To maintain precise control over temperature and pressure within the isothermal-isobaric ensemble, all MD simulations are conducted in NPT ensemble which employ two crucial techniques. The Nosé-Hoover thermostat, as introduced by Nosé \cite{Nose1984} and further refined by Hoover \cite{Kassir1985}, regulates temperature fluctuations throughout the simulations. Meanwhile, the Parrinello-Rahman barostat, developed by Parrinello and Rahman \cite{Parrinello1981}, ensures stable pressure conditions. Throughout the paper the pressure is fixed at $P = 0$. These simulations on the thermodynamic properties of our model are not meant to establish novel aspect of these materials, but rather to establish the thermodynamic and structural nature of the material as a reference pointy for understanding dynamical properties which are the novel aspect of the present work.

\begin{figure}
\centering
 {\includegraphics[width=0.45\textwidth] {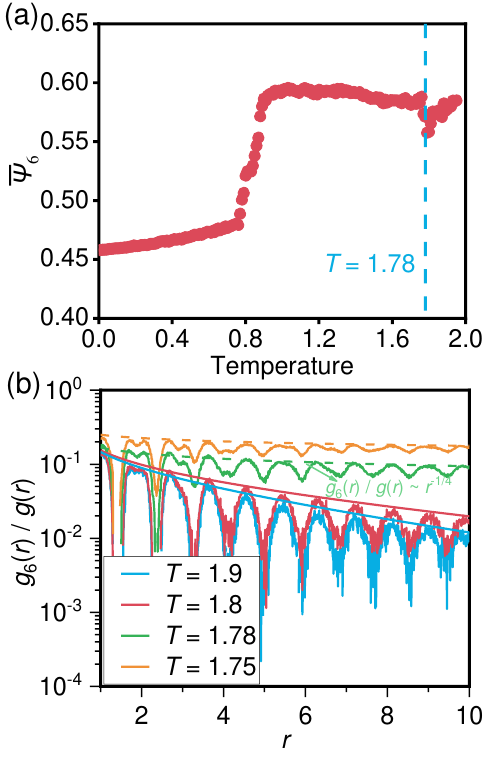}}
  \caption{(a) The temperature-dependent sixfold bond orientation order parameter. (b) Comparison of $g_6(r) / g(r)$ at various temperatures in the vicinity of $T_{LH} = 1.78$. The solid lines represent the outcomes of OZ fittings, while the dashed lines depict the results of power-law fittings. The green dashed line marks the boundary at $g_6 (r) / g (r) \thicksim r^{-1/4}$, signifying the demarcation between the liquid phase and the hexatic phase.}
  \label{fig:2}
\end{figure}

During the cooling process, the system exhibits two thermodynamic transitions. The transition from phase 1 (liquid) to phase 2 (hexatic) happens at about $T_{LH} = 1.78$. In order to quantitatively identify the location of the transition between these phases, and qualify the nature of phase 2, we calculate the six-fold bond-orientation order parameter \cite{Xu2013,Tong2018} according to the relation,
\begin{equation}\label{eq:2}
\psi_6^j = \frac{1}{N_j}\sum_{m=1}^{N_j}\exp \left( i6\theta_m^j \right),
\end{equation} 
where $N_j$ is the number of nearest neighbors of particle $j$, $\mathbf{r}_j$ is the position of particle $j$, and $\theta_m^j$ is the angle between the vector ($\mathbf{r}_m-\mathbf{r}_j$) and the $x$ axis. The cutoff distance for determining these nearest neighbors is set to be $1.2$, which corresponds to the bond length derived from the radial distribution function (RDF), $g(r)$, as discussed below. The average order parameter is then obtained via $\overline{\Psi}_6 = \frac{1}{N}\sum_{j=1}^{N} \psi_6^j$. The $T$ dependence of $\overline{\psi}_6$ is shown in Fig.~\ref{fig:2}(a), which clearly indicates two thermodynamic transitions at high and intermediate $T$. 

Furthermore, the spatial correlation of the particle-level $\psi_6^j$ can be computed using the following formula: 
\begin{equation}\label{eq:3}
g_6(r) = \frac{L^2}{2 \pi r \Delta r N \left ( N-1 \right )} \left \langle \sum_{j \neq k}^{} \delta \left( r-\left| \mathbf{r}_{jk} \right| \right) \psi_6^j \psi_6^{k \ast} \right \rangle,
\end{equation} 
where $L$ is the length of the simulation box, and $N$ the number of atoms. The quasi-long-range bond-orientation order arises from the algebraic decay of $g_6\left( r \right) / g \left( r \right) \thicksim r^{- \eta_6}$. It is noteworthy that at the boundary between the liquid phase and the hexatic phase, $\eta_6$ assumes a value of 1/4. In the liquid phase, the system exhibits only short-range bond-orientational order. Consequently, the bond-orientation correlation function is more aptly described by the Ornstein-Zernike (OZ) function, characterized by $g_6 \left( r \right) / g \left( r \right) \thicksim r^{-1/2} \exp \left( -r / \xi_6 \right)$ \cite{Kawasaki2007, Watanabe2008, Xu2013}. As evident in Fig.~\ref{fig:2}(b), $g_6 (r) / g (r)$ exhibits a more gradual decay with increasing temperature. We observe a power-law decay with an exponent of $\eta_6 = -1/4$ for $g_6 (r) / g (r)$ the characteristic temperature $T_{LH} = 1.78$, providing a practical estimate of the location of the liquid-hexatic transition \cite{Kawasaki2007, Watanabe2008, Xu2013}.
It is worth to mention a situation where quasicrystals melt without a hexatic phase. There exist situations in 2D, where quasicrystals transition into a square crystal, and then melt \cite{Engel2011}. 
While quasicrystals are often high-temperature phases, it is hypothesized that there might exist quasi-crystalline ground states, the so-called ideal tilings~\cite{PhysRevLett.53.2477,PhysRevB.34.596}.

\begin{figure}
\centering
 {\includegraphics[width=0.4\textwidth] {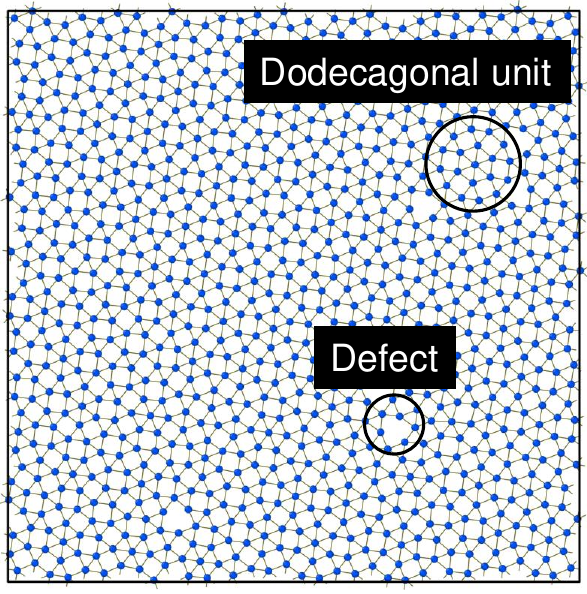}}
  \caption{The atomic structure of the monolayer dodecagonal (12-fold) quasicrystal, following relaxation at $T=0.3$. The dodecagonal unit and structural defect are highlighted by the circles.}
  \label{fig:3}
\end{figure}

The transition from phase 2 (hexatic) to phase 3 (quasicrystal) takes place near the characteristic temperature $T_{HQ} = 0.87$. Through a detailed comparison of the local configurations in phase 3 with experimental observations obtained via high-resolution transmission electron microscopy (TEM) \cite{Xiao2012}, as illustrated in Fig. \ref{fig:3}, we confirm that phase 3 corresponds to a monolayer dodecagonal (12-fold) symmetry quasicrystal, pointing to the relevance of our simulations to real-world quasicrystal material. Our model quasicrystal exhibits dodecagonal units and a point defect structure within a hexagonal ring configuration, both of which align with the observations made through TEM \cite{Xiao2012} and other experimental observation \cite{Plati2023}. Here, the number of defects can be controlled according to the cooling rate, but they are indeed unavoidable. This situation also exists in real quasicrystal materials \cite{Han2021}. The agreement between simulation results and experimental observations confirms the fidelity of our model in capturing the intricate atomic arrangement of the quasicrystal phase.
The specific quasicrystal here is the so-called square-triangle tiling (STT) in the literature \cite{Kuo1988}. Actually, it is atypical for quasicrystals, and even atypical for dodecagonal quasicrystals, due to the fact that it does not have phason modes. Instead, its dynamics is dominated by the diffusion of semi-defects (or partial dislocations) as observed in Fig. \ref{fig:3}, a process called Zippers motion \cite{Oxborrow1993}.

\begin{figure}
\centering
 {\includegraphics[width=0.35\textwidth] {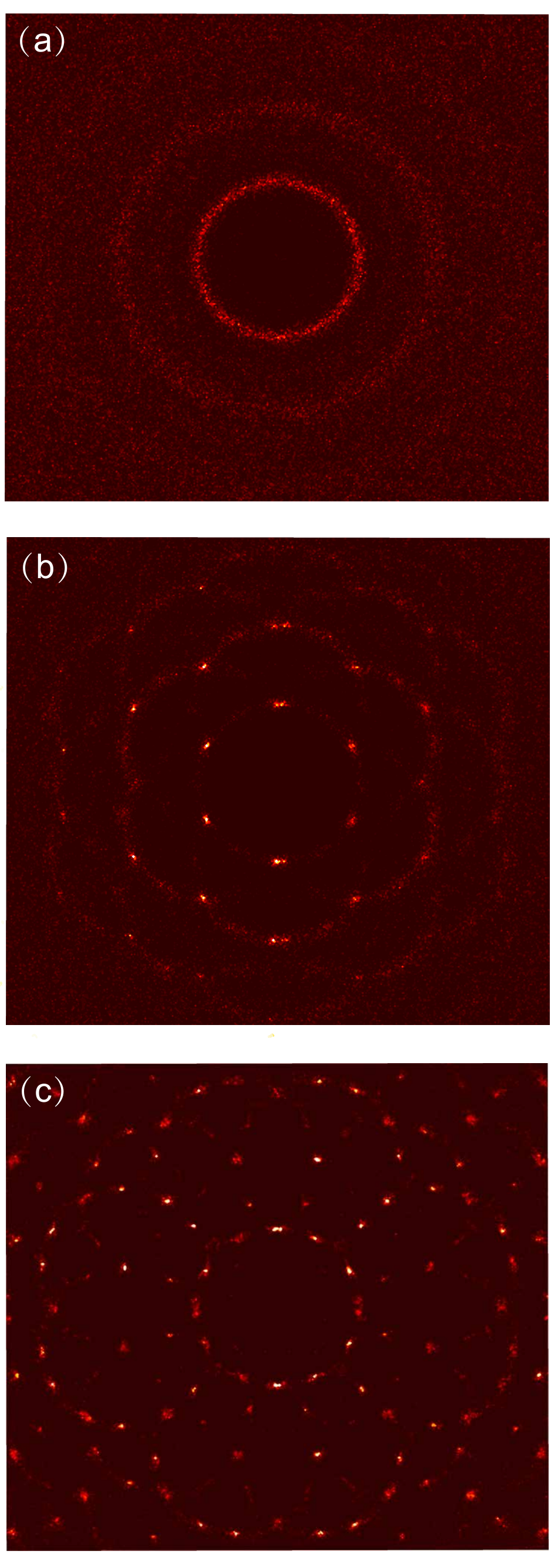}}
  \caption{Intensity map of the static structure factor of (a) liquid phase at $T = 1.95$, (b) hexatic phase at $T = 1.2$ and (c) quasicrystal phase at $T = 0.3$.
  }
  \label{fig:4}
\end{figure}

On the other hand, the static structure factor $S(k)$ is another essential parameter for describing the local structure of materials. It characterizes the system's structure in reciprocal space and can be directly measured in experiments involving elastic scattering (diffraction). In MD simulations, $S(k)$ is computed directly from atomic positions based on the relation,
\begin{equation}\label{eq:4}
S \left( k \right) = \frac{1}{N} \left \langle \sum_{j,j' = 1}^{N} \exp \left \{-i \mathbf{k} \cdot \left[ \mathbf{r}_{j}-\mathbf{r}_{j'} \right] \right \} \right \rangle.
\end{equation}
We generate the intensity maps of the static structure factor using Eq.~\eqref{eq:4}, which provide a comprehensive representation of the 2D structures included in the phase diagram. These intensity maps resemble the non-energy-resolved diffraction patterns typically obtained in diffraction experiments to determine crystal structure and symmetry. The diffraction pattern of the disordered phase (Fig.~\ref{fig:4}(a)) consists of diffuse rings of light, resulting from the lack of regular arrangement in ordered lattices. In contrast, the diffraction pattern of the hexatic phase exhibits both weak polyhedra and intense reflections with 6-fold and 12-fold symmetries (Fig.~\ref{fig:4}(b)), clearly indicating the hexatic phase intermediate nature between the liquid and solid phases. The diffraction pattern of the quasicrystal (Fig.~\ref{fig:4}(c)) is more intricate, displaying a 12-fold symmetry and well-defined reflections arranged in a quasi-periodic 2D pattern. These diffraction patterns serve to reinforce the credibility of our atomistic model. Note that for each of these diffraction patterns, we conducted parallel simulations on 50 independent configurations to enhance statistical reliability. Further quantification of the average quasicrystal structure will be the focus of our analysis in Appendix A.

We note that there have been many previous simulations of two dimensional crystalline materials in which corresponding transitions from a low temperature hexagonal crystalline state to an orientationally correlated hexatic state occurs upon heating, followed by an additional transition between the hexatic and liquid states at an even higher characteristic temperature. Although the well-known theories of these transitions developed by Kosterlitz and Thouless \cite{Kosterlitz1973} and Halperin and Nelson \cite{Halperin1978, Nelson1978} indicate that these transitions are continuous, many experimental and computational studies have established that these transitions can be either of second or first order in nature \cite{Kapfer2015}, depending on the particular form of the interparticle potential, and the presence of appreciable finite size effects can often make the assignment of transition order computationally challenging and sometimes controversial. Resolving the difference between weakly first order transitions and second order transitions can be inherently difficult \cite{Allen1983} and we avoid this technical discussion in the present work in relation to these thermodynamic transitions in quasicrystalline materials.

There have also been instructive studies of the collective dynamics of two dimensional systems exhibiting crystal melting and formation of an intermediate hexatic phase \cite{Zangi2004, VanDerMeer2015, Brinkman1982}. In the present exploratory study of a model quasicrystalline material, we simply note that the thermodynamic transitions that we observe in our quasicrystal material simulations are strikingly reminiscent to the hexagonal to hexatic and hexatic to liquid transitions established to exist in many two-dimensional crystal forming materials.
Moreover, these transitions in our quasicrystalline forming material are relatively sharp in terms of structures in both real and reciprocal space, as well as in energy and heat capacity, suggestive of a weakly first order phase transitions.
Given the limited size of our simulated materials, however, we do not insist on this designation of transition order and simply note that apparently well-defined thermodynamic transitions occur near certain characteristic temperatures at which the degree of material order and dynamics clearly occur. We are not aware of a previous work exploring these transitions in quasicrystalline materials.

Previous work on crystal melting have also found a prevalence for string-like collective motion \cite{Zangi2004, VanDerMeer2015}, a phenomenon that appears superficially very similar to the observations that we report below in our simulated model quasicrystal forming material. Despite the superficial similarities in the thermodynamic and dynamic properties of two dimensional crystal and quasicrystal forming materials, we stop short of equating their properties. Nonetheless, the evidence of evident long-range positional correlations in our quasicrystalline material point to a greater ``similarity" to crystalline materials than glass-forming materials. However, given the somewhat contradictory indications from simulation and measurements of a ``similarity" of quasicrystalline materials to both crystalline and glass-forming materials, we take a fresh look at which class of materials more closely resembles quasicrystalline materials from a dynamical perspective.

\subsection{Quasicrystal dynamics}

The mean-squared displacement (MSD) serves as a standard method to probe the dynamics of materials. The MSD normally quantifies the average displacement of a particle within the system as a function of time, expressed as,
\begin{equation}\label{eq:5}
\left \langle r^{2} \left( t \right) \right \rangle = \frac{1}{N} \left \langle \sum_{i = 1}^{N} {\lvert \mathbf{r}_i \left( t \right) - \mathbf{r}_i \left( 0 \right) \rvert}^2 \right \rangle.
\end{equation}
However, in 1D and 2D systems, long-range thermal fluctuations in positional ordering occurs, as described by the Mermin-Wagner fluctuations \cite{Mermin1966}, can pose challenges for accurately measuring the ``real" dynamics of two-dimensional materials. These fluctuations manifest as appreciable effects, as the simulation system size imposes a cutoff on the maximum wavelength of these fluctuations\cite{Illing2017, Vivek2017, Li2019}. Mitigating the finite-size effects when studying dynamics in the 2D dodecagonal quasicrystal system is especially important when using periodic boundary conditions. To address this problem, we also employ a modified method by calculating the ``cage-relative MSD'' (CR-MSD) in place of the standard MSD:
\begin{equation}\label{eq:6}
\left \langle r^{2} \left( t \right) \right \rangle _{\rm{CR}} = \frac{1}{N} \left \langle \sum_{i = 1}^{N} \left[ \left( \mathbf{r}_i \left( t \right) - \mathbf{r}_i \left( 0 \right) \right) - \frac{1}{N_i} \sum_{j = 1}^{N_i} \left( \mathbf{r}_j \left( t \right) - \mathbf{r}_j \left( 0 \right) \right) \right] ^2 \right \rangle.
\end{equation}
In this context, the second term in Eq.~(\ref{eq:6}) represents the center of mass of the cage, which is derived from the positions of the $N_i$ nearest-neighbors of the central particle $i$. This approach helps to mitigate finite-size effects and allows for a more accurate assessment of dynamics in the 2D dodecagonal quasicrystal system.

We also caution that our calculations employ periodic boundary conditions, an assumption that can greatly influence the estimated melting temperature and dynamics of crystalline materials. The use of periodic boundary conditions is adopted based on the optimist assumption that this should reduce finite size effects and give results more consistent with bulk materials of primary practical interest. In crystalline materials, this assumption can admittedly lead to difficulties because the normally higher mobility at the boundaries of the material serves to initiate melting near the thermodynamic melting temperature $T_m$. 
The apparent melting temperature of a crystalline material simulated by periodic boundary conditions is often about 20 $\%$ higher than the equilibrium melting temperature at which real materials with structure defects melt. The former corresponds to temperature at which the superheated crystal becomes unstable, the homogeneous melting temperature \cite{Zhang2013}.
We recognize that ``superheating" might also be an issue in quasicrystalline materials, but in the present exploratory study we consider the simpler case of quasicrystalline material with periodic boundary conditions.

In most cases, particles exhibit ballistic motion at very short times and eventually transition to random diffusive motion at longer timescales. As such, MSD, or the cage-relative MSD (CR-MSD in the context of a 2D system), can be employed to assess the diffusivity $D$ of a $d$-dimensional system, utilizing the Einstein relation:
\begin{equation}\label{eq:7}
D = \frac{1}{2d} \lim_{t \to \infty} \frac{\left \langle r^2 \left( t \right) \right \rangle _{\rm{CR}}}{t}.
\end{equation}
Here, $d =2$. The determination of this quantity provides an important point of contact between experiments on quasicrystalline materials and our simulations.

In addition to MSD, another observable property that quantifies dynamic relaxation is the self-intermediate scattering function (SISF), which serves to characterize density fluctuations of the same particle within the system at time $t=0$ and at a subsequent time $t$. This quantity is defined as,
\begin{equation}\label{eq:8}
F_{s}\left( k,t \right) = \frac{1}{N} \left \langle \sum_{i=1}^{N} \exp \left\{-i \mathbf{k} \cdot \Delta \mathbf{r}_i \left( t \right) \right\} \right \rangle,
\end{equation}
where $\Delta \mathbf{r}_i \left( t \right) = \mathbf{r}_i \left( t + \Delta t \right) - \mathbf{r}_i \left( t \right)$, $\Delta t$ represents a time interval. Here, the SISF is defined at the wave vector $k = 7.3$, which corresponds to the major peak in $S(k)$ and is sensitive to variations in atomic density, making it an important indicator of structural transitions, particularly in disordered phases. Besides, the temperature dependence of the first peak position in $S(k)$ is weak and the peak position is taken to be a constant. The structural relaxation time $\tau_{\alpha}$ is then defined as the time when $F_\text{s}\left( k^*,\tau_{\alpha} \right) = e^{-1}$. 

Similar to the approach used with CR-MSD, we can define a cage-relative intermediate scattering function based on Eq.~(\ref{eq:8}). Here, we subtract the motion of the center of mass of the nearest neighbors from the total displacement:
\begin{equation}\label{eq:9}
F_{s \_ \rm{CR}}\left( k^*,t \right) = \frac{1}{N} \left \langle \sum_{i=1}^{N} \exp \left\{-i \mathbf{k} \cdot \left( \Delta \mathbf{r}_i \left( t \right) - \Delta \mathbf{r}_{i \_ \rm{CR}} \left( t \right) \right) \right\} \right \rangle,
\end{equation}
where the displacement of the center of mass given by $N_i$ neighbors reads
\begin{equation}\label{eq:10}
\Delta \mathbf{r}_{i \_ \rm{CR}} \left( t \right) = \frac{1}{N_i} \sum_{j = 1}^{N_i} \left( \mathbf{r}_j \left( t + \tau \right) - \mathbf{r}_j \left( \tau \right) \right).
\end{equation}
This approach allows us to account for nearest-neighbor contributions and effectively characterize the dynamics in the presence of structural constraints.

The non-Gaussian parameter $\alpha_{2}\left( t \right)$ stands as a fundamental measure of dynamic heterogeneity in complex systems, indicating the degree of deviation from a Gaussian distribution in the displacement of all particles. In 2D systems, $\alpha_{2}\left( t \right)$ can be calculated as follows:
\begin{equation}\label{eq:11}
\alpha_{2}\left( t \right) = \frac{\left \langle \sum_{i=1}^{N} \left[ \mathbf{r}_{i} \left( t \right) - \mathbf{r}_{i} \left( 0 \right)\right]^{4} \right \rangle}{2\left \langle \sum_{i = 1}^{N} \left[\mathbf{r}_i \left( t \right) - \mathbf{r}_i \left( 0 \right)\right]^2 \right \rangle ^{2}} - 1.
\end{equation}
However, in line with the approach employed for the cage-relative MSD and SISF, we precisely calculate the cage-relative non-Gaussian parameter, which is defined as,
\begin{equation}\label{eq:12}
\alpha_{2 \_ \rm{CR}}\left( t \right) = \frac{\left \langle \sum_{i=1}^{N} \left[ \left( \mathbf{r}_{i} \left( t \right) - \mathbf{r}_{i} \left( 0 \right) \right) - \frac{1}{N_i} \sum_{j = 1}^{N_i} \left( \mathbf{r}_j \left( t \right) - \mathbf{r}_j \left( 0 \right) \right) \right]^{4} \right \rangle}{2\left \langle \sum_{i = 1}^{N} \left[ \left( \mathbf{r}_i \left( t \right) - \mathbf{r}_i \left( 0 \right) \right) - \frac{1}{N_i} \sum_{j = 1}^{N_i} \left( \mathbf{r}_j \left( t \right) - \mathbf{r}_j \left( 0 \right) \right) \right]^2 \right \rangle ^{2}} - 1.
\end{equation}
This refined approach accounts for the motion of the center of mass of the nearest neighbors and provides a more precise assessment of the non-Gaussian behavior in 2D system.

\section{Results}
\label{sec:3}

\subsection{Structural relaxation}

\begin{figure}
\centering
 {\includegraphics[width=0.45\textwidth] {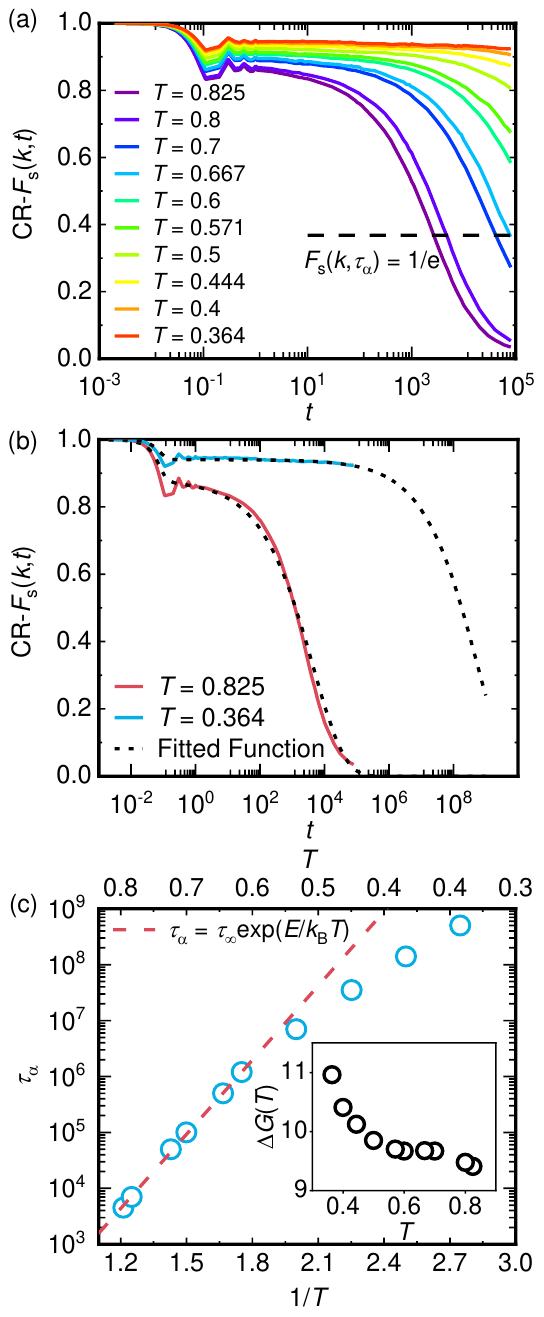}}
  \caption{Relaxation dynamics in the quasicrystal. (a) Cage-relative SISF vs time at various temperatures ranging from $T = 0.364$ to $T = 0.825$. (b) Fitted curves of the relaxation function using Eq.~\eqref{eq:13} to the cage-relative SISF data. (c) Arrhenius plot illustrating the temperature-dependent $\alpha$-relaxation time, with a dashed line indicating the linear fit obtained from high-temperature data. The inset shows the temperature dependence of the activation free energy $\Delta G(T)$.}
  \label{fig:5}
\end{figure}

We next quantify the self-intermediate scattering function and the structural relaxation or $\alpha$-relaxation time of the dodecagonal quasicrystal. Fig.~\ref{fig:5}(a) illustrates the temporal evolution of CR-$F_s(k^*,t)$ as the temperature increases, calculated using Eq.~\eqref{eq:9}. SISF is typically defined at the first peak of $S(k)$. The evolution of the SISF reveals a two-step relaxation process, including the short-time $\beta$ relaxation and the long-time $\alpha$-relaxation. The $\alpha$-relaxation time $\tau_{\alpha}$ is indicated by the horizontal dashed line in Fig.~\ref{fig:5}(a).

It is evident that structural relaxation, as measured by the self-intermediate scattering function, occurs in our quasicrystalline material, is similar to both glass-forming liquids \cite{Zhang2015} and heated crystalline solids \cite{Zhang2013} in the sense that there is a general fast beta relaxation occurring on a timescale on the order of $10^{2}$ MD steps that is relatively temperature insensitive and a longer time $\alpha$ relaxation process having a relaxation time that is highly temperature dependent.

As the temperature decreases, the $\alpha$-relaxation time increases rapidly, a well-known, even defining, attribute to glass-forming liquid, but this behavior is also characteristic of heated crystals and derives both from the reduced kinetic energy of the atoms upon cooling and the associated reduced rate of atomic diffusion \cite{Zhang2013}. Eventually, at a certain crossover temperature in the liquid regime (not studied in the present paper), the $\alpha$- and fast $\beta$-relaxation processes are expected to merge, defining an onset temperature $T_A$ for the multi-step dynamics in glass-forming, crystalline and quasicrystal materials. This pattern of relaxation thus seem to be a universal pattern of relaxation in condensed matter rather broadly. At low temperatures, the dynamics of the quasicrystal becomes so slow that it becomes increasingly difficult to make simulation observations under equilibrium conditions, a familiar situation in glass-forming liquids, but again this is the situation in heated crystalline materials \cite{Wang2021}. To quantify the observations in Fig.~\ref{fig:5}, a necessary step in comparing to the dynamics of crystalline and glass-forming materials sufficiently to enable determination which material is more like our quasicrystalline system, we fit our $F_s(k,t)$ data to a two-step relaxation function, defined as,
\begin{equation}\label{eq:13}
F_s \left( k^*,t \right) = \left( 1-A_{\alpha} \right) e^{- \left( t / \tau_{f} \right)^{\beta_{f}}} + A_{\alpha} e^{- \left( t / \tau_{\alpha} \right)^{\beta}}.
\end{equation}
This function has been found to describe $F_s \left( k^*,t \right)$ for a range of materials \cite{Zhang2017,Giuntoli2020,Cui2017prb}.

Our fitting of the fast dynamics properties $\tau_f$ and $\beta_f$ leads to $\tau_f \approx 0.1$ and $\beta_f \approx 1.9$ [Fig.~\ref{fig:5}(b)], which is rather typical of glass-forming liquids \cite{Zhang2021}. The ``non-ergodicity" parameter $A_{\alpha}$ displays relatively weak temperature dependence within the explored temperature range. Importantly, fixing these three parameters ($\tau_f$, $\beta_f$, and $A_{\alpha}$) does not significantly compromise the accuracy of our fits across the entire temperature range investigated \cite{Giuntoli2020}, allowing for precise quantification of quasicrystal relaxation.

At this stage, it is not evidently clear whether our quasicrystal material is more like a ``typical" crystalline material or a glass-forming liquid as these materials share many features in their dynamics when temperature is varied. We next consider the temperature dependence of $\tau_{\alpha}$ to search for some discriminating characteristics of the quasicrystal dynamics that could help us in this determination.

In Fig.~\ref{fig:5}(c), we present the $T$ dependence of $\tau_{\alpha}$ in our 2D dodecagonal quasicrystal in the traditional form of an Arrhenius plot. We observe a relatively high temperature regime in which $\log \tau_{\alpha}$ exhibits roughly linear variation with $1/T$, but the slope bends over sharply as the $T$ decreases below 0.5, a temperature notably in the quasicrystal phase regime. We again emphasize that we did not simulate our quasicrystalline material in the liquid regime so that there should be another Arrhenius-like regime for $\tau_{\alpha}$ at higher temperature, with probably a different slope (i.e., activation energy). We thus simulate an intermediate temperature range where multi-step relaxation of $F_s \left( k^*, t \right )$ is prevalent and thus we cannot assign the onset temperature $T_A$ corresponding to a temperature at which the $\alpha$ and fast $\beta$ relaxation processes start to merge, and where relaxation can be reliably be taken as being Arrhenius and non-cooperative to a good approximation \cite{PazminoBetancourt2018, PazminoBetancourt2015}. Qualitatively, the kink-like transition from Arrhenius to non-Arrhenius behavior aligns with numerous previous observations on some glass-forming materials \cite{Iwashita2013, Blodgett2015}, but this type of transition also occurs in also highly anharmonic crystalline materials, such as superionic crystalline materials \cite{Zhang2019}. (We will discuss this Arrhenius curve ``kink" further below in relation to its general occurrence in the intermediate forms of matter mentioned by Angell \cite{ANGELL2000791}.) Thus, we still have not arrived at the desired descrimination of quasicrystalline materials in relation to other more studied condensed materials.

Past experience on both glass-forming liquids \cite{Zhang2021} and heated crystalline materials \cite{Zhang2019} has shown the estimation of the activation energy from the local slope of this type of Arrhenius plot cannot be reliably used to define the activation energy under conditions in which the slope of the Arrhenius curve varies with $T$ can be highly misleading \cite{Zhang2021}. An estimate of the activation free energy for diffusion requires that the actual $T$ dependent activation free energy $\Delta G(T)$, formally defined by the relation, $\Delta G(T) \equiv k_{B}T \ln (\tau_{\alpha} / \tau_{0})$ \cite{PazminoBetancourt2014, 10.1063/5.0039162, Zhang2021}. Here, $\tau_{0}$ serves as a prefactor associated with the vibrational relaxation time, estimated at approximately 0.05 in our reduced units. Estimates of the temperature dependent activation energy deduced from this definition of $\Delta G(T)$ are indicated in the inset of Fig.~\ref{fig:5}(c), where we observe that $\Delta G(T)$ progressively increases upon cooling over the temperature range that we investigate. We interpret this trend to be consistent with the sigmoidal temperature variation of $\Delta G(T)$ observed in previous studies of structural relaxation of Al-Sm, a glass-forming liquid \cite{PazminoBetancourt2014, Zhang2021} that exhibits fragile-to-strong type glass-formation, and also experiments and simulation on superionic crystalline materials \cite{Zhang2019}. Again, we have not been able to resolve whether the dynamics is more like a crystal or a heated crystalline system, leading us to consider the nature of dynamic heterogeneity in the quasicrystal material in the hope of finally resolving the situation.

\subsection{A quantitative measure of dynamic heterogeneity}

To quantify ``dynamic heterogeneity", \textit{i.e.}, mobility fluctuations in the dodecagonal quasicrystal, we first consider the non-Gaussian parameter frequently used in the literature, $\alpha_2 \left( t \right)$ \cite{Andersen1995, Starr2013}, which is perhaps the simplest dynamic heterogeneity measure utilized in the study of the dynamics of condensed materials. This dynamical property quantifies the degree of deviation of particle displacements from a Gaussian distribution and this phenomenon has often been found in association with large mobility fluctuations in complex liquids, making this quantity an indirect measure of such dynamic heterogeneity.

\begin{figure}
\centering
 {\includegraphics[width=0.45\textwidth] {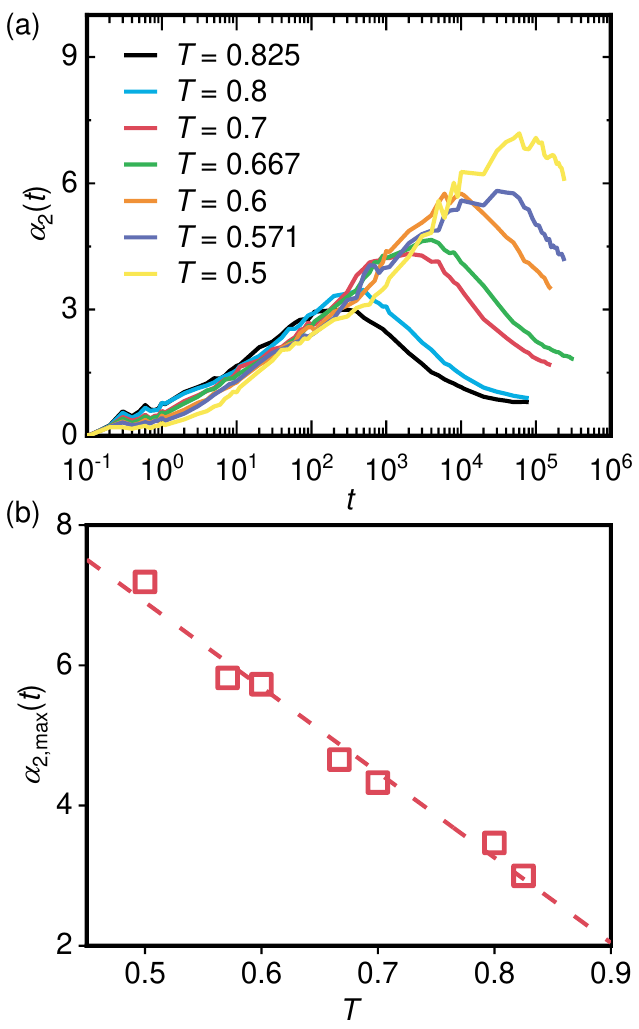}}
  \caption{(a) Cage-relative non-Gaussian parameter $\alpha_2$ as a function of time at different temperatures. (b) Temperature-dependence of the maximum value of the CR-$\alpha_2$, \textit{i.e.}, $\alpha_{2,\text{max}}$. We expect the linear variation in this figure to only hold over a limited $T$ range.
  }
  \label{fig:6}
\end{figure}

The time evolution of CR-$\alpha_2 \left( t \right)$ for the dodecagonal quasicrystal is depicted in Fig.~\ref{fig:6}(a) at various temperatures. Similar to glass-forming liquids, $\alpha_2(t)$ exhibits a peak at a characteristic time $t^*$ \cite{Andersen1995} that is generally shorter than $\tau_{\alpha}$. This measure of the ``lifetime" of the dynamic heterogeneity increases upon decreasing temperature, a trend typical of glass-forming liquids, but contrary to heated crystals where this quantity appears to increase upon heating from the low temperature crystal state where this quantity is very small, but then passes through a maximum at intermediate temperatures and then falls upon approaching the melting temperature \cite{Zhang2013}. This decreases of $\alpha_2$ upon cooling was found to be especially prevalent in the simulation of superionic ${\rm UO}_{2}$ in its high temperature \cite{Zhang2019}. This is the first strong hint that the dynamics of quasicrystalline material is perhaps more like a glass-forming liquid at low $T$ than a heated crystal. However, we do not consider the $T$ regime at which the hexatic phase of the quasicrystalline material enters the liquid states, the analog of the melting in the three-dimensional crystalline material studied previously. In contrast, we find that $\alpha_{2,\text{max}}$ progressively increases upon lowering $T$, a pattern of behavior that aligns with the dynamics of glass-forming liquids rather than crystalline materials \cite{Kob1997}. This trend also clearly points to a commonality between metallic glass-forming liquids and our quasicrystalline material. We next consider the interrelation between diffusion and the structural relaxation time and $t^*$, which as we shall see offers further and more compelling evidence for a closer relationship between the dynamics of our quasicrystalline material and glass-forming material rather than the dynamics of heated crystalline materials.

\subsection{Revisiting the kink in the Arrhenius curve for particle diffusion}

\begin{figure}
\centering
 {\includegraphics[width=0.45\textwidth] {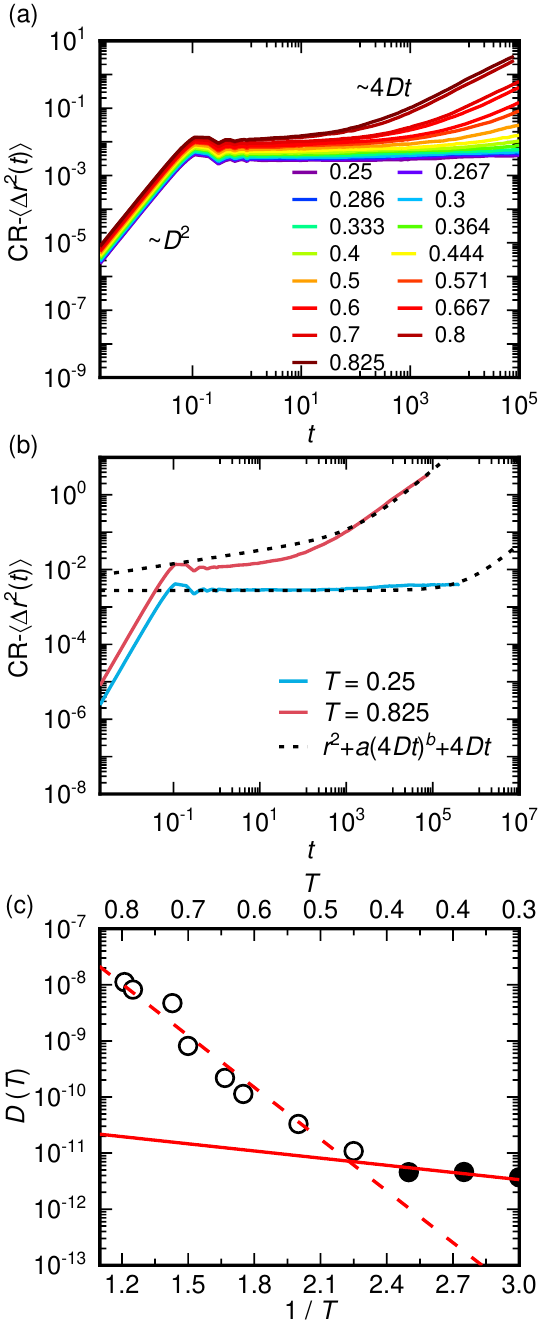}}
  \caption{(a) Cage-relative MSD at different temperatures. (b) Fit of MSD to the von Schweidler expression at two representative temperatures. (c) Arrhenius plot of diffusivity, where a striking transition of mechanism is found at $T = 0.44$. At high temperatures (hollow points), $D_{0} = 1.07 \times 10^{-4}$ and $\Delta E = 7.38$, while for low-temperature regime (solid points), $D_{0} = 6.07 \times 10^{-10}$ and $\Delta E = 1.54$.
  }
  \label{fig:7}
\end{figure}

In Fig.~\ref{fig:7}(a), we show the evolution of the temperature dependence of CR-MSD for the dodecagonal quasicrystal when the temperature is varied over a large range. As found in condensed materials rather generally, particle MSD exhibits three stages that mirrors these same dynamical regimes in $F_s(k^*,t)$. At short times, particles undergo ballistic motions, \textit{i.e.}, a particle does not collide with any particles at very short times so that the motion resembles particles in a gas. During this stage, the distance traveled varies nearly linearly with time, resulting in MSD increases proportionally to $t^2$. Beyond this ``inertial dynamics" regime, a plateau in $\left\langle\Delta r^2 \left(t\right)\right\rangle$ develops, which becomes more pronounced at lower temperatures. This pattern of behavior is characteristic of condensed materials rather broadly so it is no surprise that we see the same basic trend in our quasicrystal material. At long times, the particles escape from their ``cages" at intermediate time in which most of the atoms vibrate within regions defined by surrounding particles. The Debye-Waller parameter $\left\langle u^2 \right\rangle$ describes the amplitude of atomic vibrations on the fast $\beta$ relaxation time [see Fig.~\ref{fig:5}(a)], defines the onset time of caging and the ``end" of the fast $\beta$ relaxation process. At sufficiently long times, particle motion can be described as simple diffusion, provided the material is in thermal equilibrium. The timescale required for this long time diffusion emerges, corresponding to a time on the order of $\tau_{\alpha}$ can be extremely large and difficult to access by molecular dynamics simulation. This can make the estimation of the diffusion derived from particle displacement rather difficult to estimate computationally as we shall next show.

As in glass-forming, we utilize the semi-empirical expression \cite{Andersen1995, Schroder2020} to estimate $D$ from the widely utilized relation on this context,
\begin{equation}\label{eq:14}
\left \langle \Delta r^2 \left( t \right) \right \rangle = r_0^2 + a \left( 4Dt \right)^b + 4Dt.
\end{equation}
While this method of estimating $D$ provides a reasonable method for estimating $D$ in our quasicrystalline material, the slowing of the dynamics limits the temperature range in which estimates can be made and we confine our estimates to the same $T$ range as we considered above for $\tau_{\alpha}$. Fig.~\ref{fig:7}(b) shows the result of our fitting procedure based on Eq.~\eqref{eq:14} to estimate $D$ as function of $T$.

As in the case of $\tau_{\alpha}$, we see that $D$ exhibits two distinct temperature regimes when the data is considered in the form of an Arrhenius plot, as depicted in Fig.~\ref{fig:7}(c). Measurements of $D$ in a real quasicrystalline material have indicated a striking similar ``kink" in the Arrhenius curve for $D$ \cite{PhysRevLett.80.1014}, confirming this interesting dynamic transition found in our simulations. Recent simulations of $D$ for a model metallic glass material \cite{Zhang2021} also strikingly resemble our observations in Fig.~\ref{fig:7}(c), where a kink change arises in the Arrhenius curve for $D$ data indicating a clear change in the dynamical regimes of the material where the origin of this change is investigated in depth \cite{Zhang2021, 10.1063/5.0039162}. Arrhenius curves having a ``kink" as in Fig.~\ref{fig:7}(c) having upward curvature characteristically arise in the dynamics of enzymes \cite{Truhlar2001, Nagel2011, Liang2004}, an observation in line with Angell's suggestion that globular proteins could also be viewed as an intermediate form of matter having properties in common with other materials of this type. It was mentioned before that this type of ``kink" in the Arrhenius curves is found in Al-Sm metallic glass-forming liquids \cite{Zhang2021, 10.1063/5.0039162} and, moreover, this type of non-Arrhenius dynamics is observed in simulations and measurements of the dynamics of model superionic materials, and the interfacial dynamics of crystalline materials \cite{Manley2019}, providing further evidence supporting Angell heuristic arguments about the existence of ``intermediate" forms of matter to crystalline or liquid states. Below, we show that the temperature variation of $D$ and $\tau_{\alpha}$ bears a direct relation to $t^*$ and $\tau_{\alpha}$ so we do not provide a discussion of the $T$ dependent activation energy as it must be the same as deduced from Fig.~\ref{fig:7}(c).

\subsubsection{Atomic-scale mechanisms governing relaxation and diffusion in the fast $\beta$ and $\alpha$ relaxation regimes}

\begin{figure*}
\centering
 {\includegraphics[width=1\textwidth] {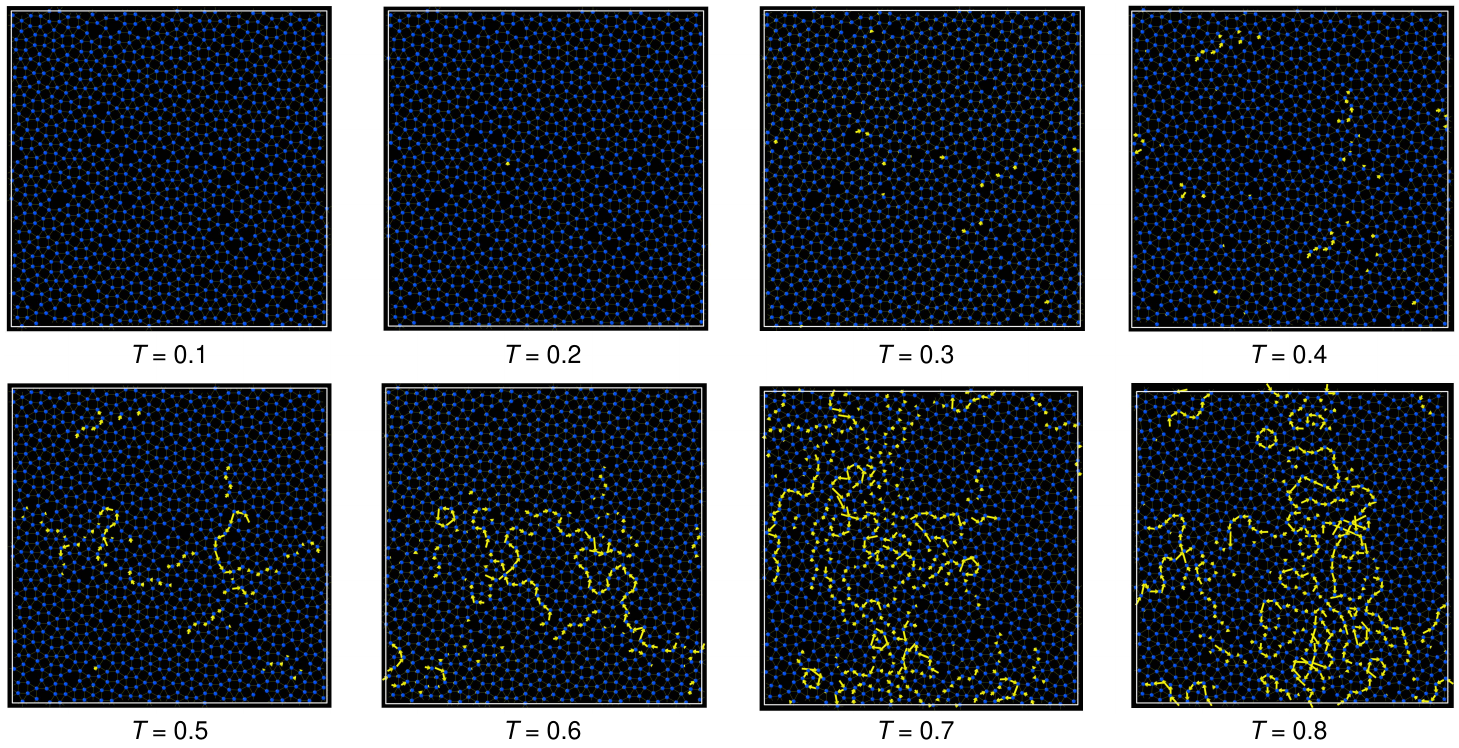}}
  \caption{Characterization of the atomic motion mechanisms in quasicrystal by displacement vector (yellow arrows) at different temperatures. The emergence of collective 1D string-like structures is evident at high temperatures.
A supplementary video 2 is provided to visualize the dynamic process of string-like motion.
  }
  \label{fig:8}
\end{figure*}

In this section, we investigate the mechanisms of the transition between a distinct fast $\beta$ dynamics regime and $\alpha$ relaxation and the curious kinks in the Arrhenius curves for $\tau_{\alpha}$ and $D$, based on a direct examination on the shape of particle trajectories and the occurrence of collective motion in which more than one particle is involved. First, we examine the atomic trajectories and correlated motion at different temperatures over $3 \times 10^7$ MD steps, as shown in Fig. \ref{fig:8}. These ``defects" of a dynamical nature are widely believed to transform energetically comparable configurations into others and remind us of moving interstitials in crystalline materials. Larger scale collective motions might possibly occur at these low temperatures, but, if they do, then they must occur rarely so that they cannot be convincingly sampled by MD simulation. Starting about $T = 0.5$, string-like particle exchange motion or ``string-like" motion clearly emerges regardless of whether these dynamic structures are related to any unique mode in quasicrystal. In particular, this form of collective motion involving a permutational exchange motion starts to become conspicuous. Previous MD simulations on both crystalline materials and glass-forming liquids exhibit what appears to be a superficially a geometrically similar collective motion so this type of motion to be rather universal in condensed materials at finite temperatures. Importantly, the $T$ regimes associated with these distinct regimes of particle diffusion and relaxation align with the kink observed in Fig.~\ref{fig:5}(c) and Fig.~\ref{fig:7}(c).

\begin{figure}
\centering
 {\includegraphics[width=0.5\textwidth] {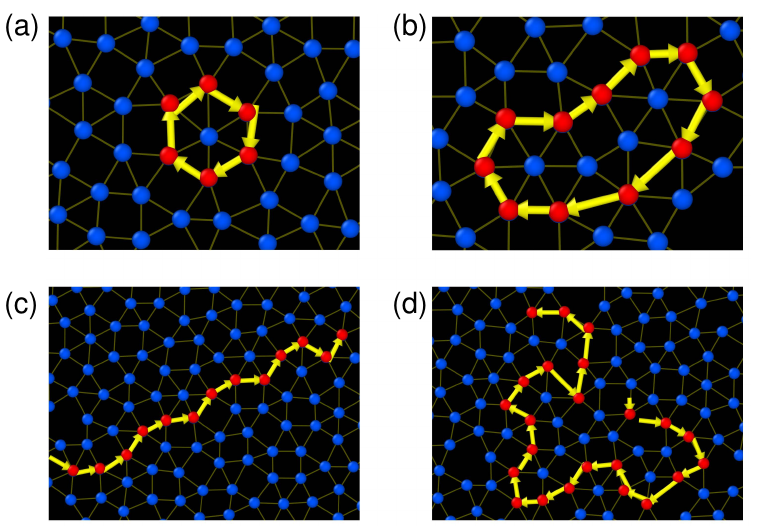}}
  \caption{Typical string-like motion mechanisms at high temperature $T = 0.8$. (a) A single loop. (b) A large loop containing several hexagonal units. (c) A long straight line. (d) A serpentine line. The yellow arrows represent displacement vectors.
  }
  \label{fig:9}
\end{figure}

At noted previously, diffusion becomes conspicuously dominated by particle exchange motion at higher $T$. While string-like motions have been extensively studied in glass-forming liquids and heated crystalline materials \cite{PhysRevLett.80.2338, Wang2018, McKenzie-Smith2022}, such observations are rarely reported in quasicrystalline systems \cite{PhysRevB.82.134206, Lieu2022}. And no previous work has studied string-like collective motion in quasi-crystals in any quantitative way, nor has any previous work on quasi-crystals related the string properties to the dynamics following analyses given previously in the study of glass-forming liquids \cite{Zhang2019, Xu2013, PhysRevLett.80.2338, Starr2011}. Fig.~\ref{fig:9} illustrates typical string-like configurations at $T = 0.8$. Notably, string-like motions in quasicrystals exhibit a rich diversity of forms, encompassing both closed and open strings. Figs.~\ref{fig:9}(a) and \ref{fig:9}(b) show the two primary forms of these dynamic structures: closed strings or ``rings" and open linear strings. Interestingly, particles in the inner cluster appear to be more active than those residing outside, indicating a possible correlation between the soft spots and the strings. As for open strings, they may manifest as either highly persistent even linear structures (Fig.~\ref{fig:9}(c)) or coil-like random structures reminiscent of flexible polymer chains (Fig.~\ref{fig:9}(d)). The whole dynamic process is shown in Supplemental Video 2. Upon quantification, we find that ring-like form of the strings account for 8 $\%$ to 15 $\%$ of the total string-like structures, with this ratio increasing as the system is cooled down, a phenomenon observed previously in metallic glass-forming liquids \cite{Zhang2013}. The specific method for the determination of the strings and their statistically defined properties is described below.

\subsection{Dynamics of string-like cooperative motion and its relation to the activation energy of structural relaxation and diffusion}

In recent years, extensive atomistic simulations have unveiled mechanisms of string-like cooperative motion in amorphous solids, particularly in metallic glasses (MGs), occurring at elevated temperatures around the glass transition or above \cite{Zhang2021, Yu2017, Yang2022, Chang2022}. This phenomenon has emerged as a characteristic feature of glass-forming liquids generally. In our study, we extend this type of investigations to quantify such motions in the diffusion of the 2D dodecagonal quasicrystal and the relation between these motions to the $T$ dependent activation energy. To initiate our exploration, we provide a concise overview of the methodology employed to quantify such string-like cooperative motion.

The initial step involves identifying the ``mobile" particles that facilitate the dynamics of the condensed material. Here, mobile particles are defined as those exhibiting a displacement magnitude $r(t)$ greater than the typical amplitude of particle vibration over a specified time interval $t$. Observations of the diffusion mechanism indicate that particles engaged in the string-like motion typically undergo a multi-stage process to complete a collective motion event so that assigning a definite activation energy from an analysis of individual stages in this process is difficult. However, the largest $\Delta E$ occur at lower $T$, a trend that parallels the growth of the activation energy upon cooling apparent in the insert to Fig.~\ref{fig:5}(c).

Next, we employ a mathematical criterion that has been applied in numerous previous studies of glass-forming liquids and heated crystals to identify these mobile particles based on a threshold of atomic displacement, specifically we apply the commonly invoked condition $\left| \mathbf{r}_{i}(\Delta t)-\mathbf{r}_{i}(0) \right| > 0.6$. In this case, the selected particles have climbed over the energy barrier. Importantly, the count of mobile particles is inherently temperature-dependent, decreasing as the system cools. To identify correlated atomic motion, we employ an established procedure outlined in Ref.~\cite{Donati1998}. Collective motion implies that the spatial relationship between two particles is retained to some extent as they traverse the system. Specifically, for a pair of ``mobile" particles, denoted as particle $i$ and particle $j$, we consider them to be engaged in string-like motion if they persist within proximity to each other  over time. This proximity relation is defined by the condition $\left| \mathbf{r}_{i}(\Delta t)-\mathbf{r}_{j}(0) \right| < \delta$, where $\delta$ is a parameter chosen to be at least smaller than half of the bond length. In our case, we set $\delta = 0.2$. Importantly, we confirm that the results of string-like motion are not significantly influenced by the choice of $\delta$ in the reasonable range.

\begin{figure}
\centering
 {\includegraphics[width=0.45\textwidth] {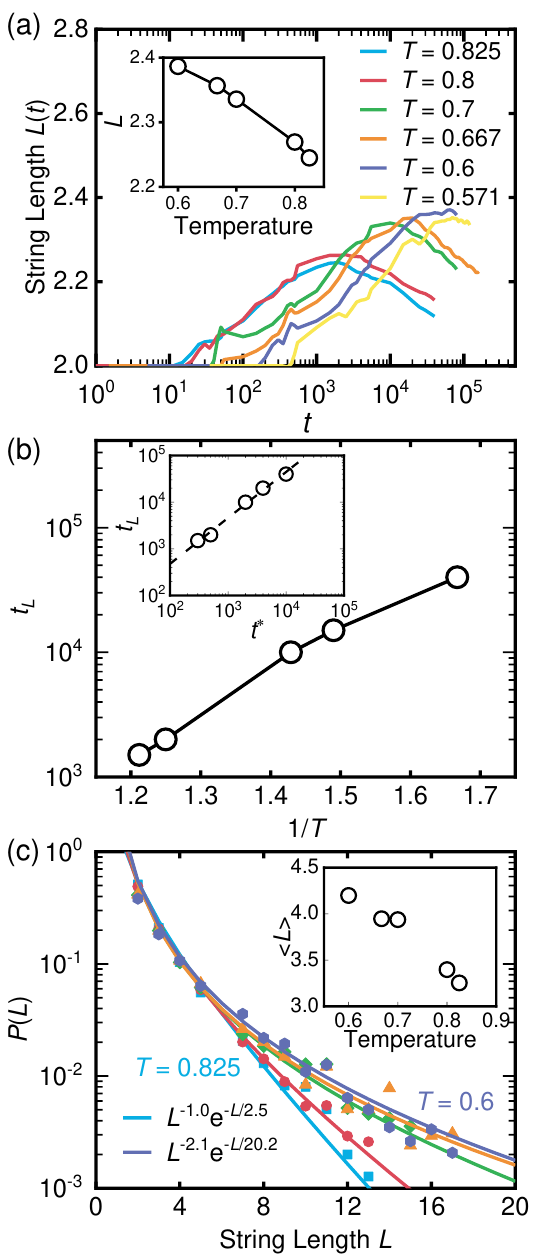}}
  \caption{Analysis of the string-like cooperative motions. (a) Average string length $L(t)$ as a function of time at different temperatures. The inset shows the $T$-dependence of the peak value. (b) Diffusive relaxation time $t_L$ vs $1/T$. The inset shows a linear scaling between $t_L$ and $t^*$. (c) The distribution of string lengths $P(L)$ at different temperatures, fitted by an exponential law. The inset exhibits the average length of the strings versus temperature. }
  \label{fig:10}
\end{figure}

Utilizing the methodologies outlined earlier, we extract several crucial parameters, including the number of mobile particles participating in string-like motion and the fraction of strings. In Fig.~\ref{fig:10}(a), we illustrate the average string length (number of particles) at various temperatures. In alignment with previous observations \cite{Donati1998, Starr2011}, the average length $L(t)$ is negligible within a very short time (ballistic motion) and progressively increases with time intervals after long-range diffusion occurs, signifying weak correlation of mobile strings. Notably, $L(t)$ exhibits a peak at a characteristic time denoted as $t_L$. This timescale defines the characteristic string ``lifetime" \cite{PazminoBetancourt2014}. The peak value intensifies with cooling, indicating enhanced cooperative mobility at lower temperatures. Simultaneously, the position of the peak shifts to longer times as the temperature decreases, suggesting a deceleration in cooperative string-like diffusion.

Upon further examination, it becomes evident that the string lifetime is approximately proportional to $t^*$ (inset of Fig.~\ref{fig:10}(b)), where the latter represents the peak time of the non-Gaussian parameter (Fig.~\ref{fig:6}(a)). This correlation is consistent with previous observations in other systems \cite{Starr2013, Zhang2011, Zhang2021, Donati1998}. Earlier studies have already established a connection between $t^*$ and $D$, i.e., $D/T \sim (t^*)^x$, as we discuss below. This linear scaling relation is contrasted by the common appearance of a non-linear relation between $D/T$ and $t^*$ with $\tau_{\alpha}$ that is generally observed in glass-forming liquids that is often referred to as ``decoupling''. Specifically, it is widely observed that the structural relaxation time experiences a more rapid increase with decreasing temperature compared to the time scale associated with the diffusion coefficient $D$ that leads to a breakdown of the Stokes-Einstein relation, $D/T \sim (\tau_{\alpha})^y$ observed in glass-forming liquids above the onset temperature for non-Arrhenius dynamics. We will provide further support for this decoupling relation below where observe a similar behavior as observed in MG forming materials.

Additionally, the characteristic time $t_L$ is slightly longer than the characteristic time $t^*$, as illustrated in the inset of Fig.~\ref{fig:10}(b). Moreover, the time scale of $t_L$ is inherently smaller than $\tau_{\alpha}$ due to the breakdown of the Stokes-Einstein relation \cite{Starr2013}. The diffusive relaxation times $t_L$ and $t^*$, occupy an intermediate position between the fast $\beta$ relaxation time and the slow $\alpha$ structural relaxation time. 

Finally, to quantify the geometrical properties of the strings, we analyzed the distribution of string length, denoted as $P(L)$, at different temperatures at the characteristic time scale $t_L$. The accurate length calculations involved a meticulous three-step process: (1) Filtering particles engaged in string-like motions. (2) Identifying the onset particle of each string from the string particles, where the onset particle does not overlap with the positions of other string particles over the time interval $\Delta t$ of observation. (3) Sequentially selecting string particles starting from the onset particle, allowing us to determine not only the length of each string, but also its topology (i.e., whether it is a closed or open string).

As depicted in Fig.~\ref{fig:10}(c), the distribution functions $P(L)$ at different temperatures exhibit an approximately exponential decay, as noted in previous work \cite{PazminoBetancourt2014,Starr2013,Donati1998}. The deviations form exponential behavior at different $T$ can be described by a power-law variation with an exponential cut-off,
\begin{equation}\label{eq:15}
P\left(L\right) \sim L^{-\theta}\exp\left(-L/L_{0}\right)
\end{equation}
where $L_0$ is a length parameter proportional to the average string length $\langle L \rangle$. This type of power law times an exponential scaling for the distribution of string length has been validated across various systems, including glassing-forming liquids \cite{PazminoBetancourt2014}, attractive colloidal particles \cite{Rouwhorst2020}, and lipid membranes \cite{Dudowicz1999, Starr2014}, where $\theta$ exhibits a similar value. Here, we find the exponent $\theta$ increases on cooling, which may reflect a progressive topological transition between open strings at high temperature to gradually increasing closed clusters at low temperature, as observed in our system and previous studies \cite{PazminoBetancourt2014, Zhang2013, VanWorkum2005}.

The inset of Fig.~\ref{fig:10}(c) depicts the temperature dependence of the average string length $\langle L \rangle$. Clearly, the size or average length of the strings increases as $T$ decreases, qualitative accord with the $T$ dependence of $\Delta G$ in Fig.~\ref{fig:5}(c). We suggest the noise in our $L(T)$ data reflects the uncertainty created by the $T$ dependence of the $\theta$ exponent in Eq.~(\ref{eq:15}). In order to obtain a more accurate equation for $P(L)$, much more quantitative estimates need to be performed on $L(t)$. Our estimates of the $T$ dependence of the activation energy are also a bit noisy and the estimation of this $T$ dependence needs further attention. Qualitatively, however, the growth of the activation free energy upon cooling is accompanied by the scale of collective motion, a signature feature of glass-forming liquids.

\subsection{Interrelation between $t^*$, $\tau_{\alpha}$ and $D$}

The growth in the peak height of $\alpha_2(t)$ (Fig.~\ref{fig:6}) reflects the strongly dynamic heterogeneity in the quasicrystal on cooling. As in glass-forming liquids, this quantity exhibits a peak at characteristic time $t^*$. The peak value of $\alpha_2\left(t^*\right)$ increases as the system progressively cooled. As noted before, studies have established a robust correlation between $t^*$ and $D$ in both atomic glass-forming liquids and heated crystals \cite{Zhang2019, Douglas2016, Zhang2015, Starr2013, Zhang2013, Zhang2015-1}. In the following discussion, we also explore whether power-law scaling relation between $D / T$ and the structural relaxation time $\tau_{\alpha}$ or a ``decoupling relation" exists in our quasicrystal material.

\begin{figure}
\centering
 {\includegraphics[width=0.45\textwidth] {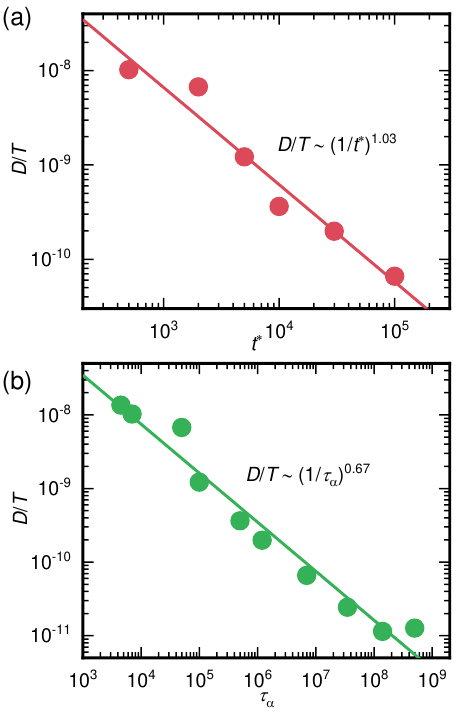}}
  \caption{Scaling of the temperature normalized self-diffusion coefficient $D$ with (a) the non-Gaussian characteristic time $t^*$ and (b) the $\alpha$ relaxation time $\tau_{\alpha}$.
  }
  \label{fig:11}
\end{figure}

We next consider the temperature dependence of the self-diffusion coefficients obtained from the long time limit of the MSD, with $t^*$ derived from the variation of the non-Gaussian parameter in $\alpha_2(t)$ (see Fig.~\ref{fig:11}(a)). This comparison reveals a robust power-law relationship between $D / T$ and $t^*$, expressed numerically as $D/T \sim (1/t^{*})^{1.03}$. Notably, the scaling exponent is equal to unity to within numerical uncertainty. This scaling evidently resembles the dynamics of metallic glass-forming liquids modeled by embedded atom method (EAM) potentials \cite{Andersen1995, Zhang2013, Starr2013} and the Kob-Anderson model in which a simple binary Lennard-Jones fluid is often studied as ``generic" glass-forming liquid \cite{Andersen1995}. Notably, this inverse scaling relation was found to not hold in simulations of superionic crystalline ${\rm UO}_{2}$ \cite{Zhang2019}, pointing again to a commonality of the dynamics of our quasicrystalline system with the dynamics of glass-forming liquids. Moreover, our finding establishes the new notion that $t^*$ can serve as a distinctive characteristic timescale for diffusion in the quasicrystalline system.

We next explore the quantitative relationship between the characteristic timescale of diffusion $t^*$ just established and the structural relaxation time $\tau_{\alpha}$. Many previous studies of glass-forming liquids have indicated that $t^*$ is related to $\tau_{\alpha}$ by a apparently universal scaling relation \cite{Zhang2015}: $t^* \sim ({\tau_{\alpha}})^{1-\zeta}$. Here, $\zeta$ defines the ``decoupling exponent", quantifying breakdown of the Stokes-Einstein relationship in glass-forming liquids given that $D/T$ scales as $1/t^*$. The case in which $\zeta=0$ corresponds to a situation in which the classical Stokes-Einstein relation applies. This ``decoupling" phenomenon is a signature feature of metallic glass-forming liquids \cite{Zhang2015, Douglas2016} and glass-forming liquids more broadly \cite{Douglas1998}, but no decoupling is observed in crystalline materials \cite{Zhang2013, Zhang2019, Zhang2015-1}. This provides a sharply defined criterion for determining whether the relaxation dynamics of quasicrystalline materials is more like crystalline or glass-forming materials.

Finally, we consider the ``fractional Stokes-Einstein relation" in the dodecagonal quasicrystal system. Figure~\ref{fig:11}(b) shows this scaling relationship holds between $D/T$ and $\tau_{\alpha}$, where we see the specific scaling relation, $D/T \sim (1/\tau_{\alpha})^{0.67}$. The decoupling exponent $\zeta$, for our quasicrystal material, is approximately 0.33 which is rather typical value for glass-forming liquids \cite{Douglas1998, Zhang2015, Douglas2016}. More generally, $\zeta$ is found to lie in the range 0.16--0.46. Specifically, previous studies have estimated $\zeta$ as 0.26 in the case of a Pd-Cu metallic glass-forming alloy \cite{Richert2007}, and 0.18 and 0.17 was reported for the Kob-Anderson binary LJ model in three dimensions \cite{Eaves2009, Sengupta2013}. A breakdown of the Stokes-Einstein relation or the ``decoupling phenomenon'' is clearly observed in our quasicrystalline system, a characteristic of glass-forming liquids generally. In contrast, simulations of heated crystalline systems \cite{Zhang2013, Zhang2019, Zhang2015-1} have consistently shown the lack of decoupling (i.e., $\zeta = 0$) to within numerical uncertainty, even these systems show a peak in the non-Gaussian parameter, strings and other features of the dynamics of glass-forming liquids. We have finally been able to find definitive simulation evidence indicating that quasicrystalline materials being more like metallic glass-forming systems in terms of their dynamics. On the other hand, we may expect the collective intermediate scattering function to not decay at long times as in crystalline materiasl \cite{Zhang2013} because of the solid nature of quasicrystal materials so that the unequivocal conclusion that quasicrystals are exactly like glass-forming liquids in terms of their dynamics does not quite hold. These materials apparently exist in a kind of material state "Twilight Zone", in which some properties bear some resemblance to materials of common experience, while others do not conform.

\section{Conclusion}
\label{sec:4}

In summary, our investigation has comprehensively explored the thermodynamic and dynamic attributes of one-component dodecagonal quasicrystals in two dimensions, employing the isotropic LJG pair potential as a benchmark model system. The thermodynamic stability of the dodecagonal quasicrystal is attained under well-defined parameter conditions. Notably, the transition from the liquid phase to the quasicrystal phase is marked by two distinct sharp thermodynamic transitions, although the actual order of these transitions remains to be determined. The consistent alignment of our model system with previous experimental findings supports the generality of our conclusions.

Our examination of the thermodynamic properties and structure entails a comprehensive analysis of density, heat capacity, specific heat, radial distribution function, and static structure factor within the prepared quasicrystal. The quasicrystal exhibits structural characteristics reminiscent of crystalline solids. The RDF of the quasicrystal unveils both short-range order and, to some extent, questionable long-range order, which is due to the assumptions inherent to the model of the LJG pair potential.

We generally infer that the dynamic properties of quasicrystals more closely parallel those of metallic glass-forming than crystalline materials, while their structures appears to have a closer relation to crystals, as inferred from thermodynamic properties relating to the existence of ordered states having translational and rotational symmetries. Quasicrystals evidently really are a hybrid form of matter. The temperature dependence of the structural relaxation time $\tau_{\alpha}$ and $D$ both exhibit a transition in the quasicrystal regime characterized by kink in Arrhenius curves for $\tau_{\alpha}$ and the diffusion coefficient having a singular upward curvature, a phenomenon observed in rather generally phenomenologically in materials that Angell has classified as intermediate forms of matter \cite{ANGELL2000791}.

As the quasicrystal material is cooled, we observe an increase in dynamic heterogeneity paralleling the slowing down of the dynamics. The characteristic time of diffusion coefficient, $t^*$, at which the non-Gaussian parameter exhibits a maximum, reflecting the magnitude of the dynamic heterogeneity in the system. Notably, the characteristic time $t^*$ is smaller than the structural relaxation time $\tau_{\alpha}$. Our comprehensive analysis of the self-diffusion of quasicrystals takes into account minimum energy pathways, revealing the intricate pathways and intermediate features of string-like motions within the dodecagonal quasicrystals. Although heated crystalline materials and glass-forming liquids exhibit many similar features in their dynamical properties, in the final analysis the dynamic properties of quasicrystals show a greater resemblance to metallic glass-forming materials than heated crystals. Previous neutron scattering studies \cite{PhysRevLett.59.102} have repeatedly arrived at the same conclusion based on notably different dynamical properties \cite{ANGELL2000791}. Measurements on $D$ in real quasicrystal materials have also indicated a kink in their Arrhenius curves for the diffusion coefficient \cite{PhysRevLett.80.1014}, as we observe in our simulations of diffusion in our quasicrystalline material. Our simulations are thus broadly consistent with measurement studies on quasicrystalline materials.
Note that heterogenous dynamics typically reflect a continuous distribution of activation energies and jump distances as that in glasses, indicative of a complex energy landscape~\cite{Schroder2020}. By contrast, in quasicrystals, the rearrangements are discrete, with only a finite number of distinguishable local configurations. In this sense, the dynamics in quasicrystals more closely resemble defect-mediated diffusion in crystals (e.g., vacancy or dislocation motion).

Another striking aspect of the dynamics of our quasicrystalline material, shared with glass-forming materials broadly, is that the activation free energy and the string length both increase upon cooling. Our data for both of these quantities exhibit some scattering and a greater effort needs to be made to estimate these properties with less uncertainty. Nonetheless, there is a clear evidence that the activation barrier for particle displacement to a distance required for diffusion and structural relaxation clearly grows upon cooling in our quasicrystal materials and glass-forming liquids generally. This is apparently a new finding that we did not ourselves initially anticipate.

Over an extended observation period, the string length at different temperatures follows a length distribution obeying a power law times an exponential scaling.
The characteristic time which the string length peaks, $t_L$, exhibits a strong correlation with diffusive relaxation time. Additionally, the present quasicrystal exhibits a breakdown of the Stokes-Einstein relationship between $D$ and $\tau_{\alpha}$, akin to many previous observations on glass-forming materials, but not to crystalline materials where this decoupling phenomenon is not observed. Of course, it is not clear why the Stokes-Einstein relation should apply to either quasicrystalline or crystalline materials, which are solids rather than liquids. 

Further research is imperative to enhance our understanding of the phason flip motions in more typical quasicrystals and how these motions fit in with the conspicuous string-like collective dynamics, and its interrelation with vibrational properties. In particular, the present understanding of the vibrational properties of quasicrystals is incomplete in our view. For example, the potential existence of an excess vibrational mode, akin to the boson peak observed in amorphous solids and heated crystals \cite{ramos2022low}, poses intriguing questions about these non-phonon modes. The possibility that the modes predicted to be responsible for phason flips might involve such excess modes merits a thorough computational investigation in itself. Unraveling the intricacies of these vibrational modes in quasicrystals and the significance of these modes for the properties of these materials will undoubtedly contribute to a broader understanding of the dynamics of quasicrystalline materials. However, we think the pursuit of these questions should be based on a quasicrystal model in three dimensions and we plan to study this type of material to answer some of remaining questions about the dynamics of the quasicrystalline materials.


\section*{Supplementary Material}
Two supplementary movies are provided to visualize the atomistic mechanism of defect hopping and string-like motions.

\section*{Acknowledgement}
This work was financially supported by the National Natural Science Foundation of China (Grant No. 12472112), and the
Strategic Priority Research Program of Chinese Academy of Sciences (Grants No. XDB0620103 and No. XDB0510301). M.B. acknowledges the support of the Shanghai Municipal Science and Technology Major Project (Grant No. 2019SHZDZX01) and the sponsorship from the Yangyang Development Fund. 
W.-S.X. acknowledges the support from the National Natural Science Foundation of China (Grant Nos. 22222307 and 21973089).


\appendix

\label{appendix}

\section*{Appendix A: Thermodynamics of dodecagonal quasicrystal}
\label{Appendix:A}
\setcounter{equation}{0}
\renewcommand{\theequation}{A\arabic{equation}}

In this section, we consider an exploration of the thermodynamics of the LJG system including the dodecagonal quasicrystal during cooling.

\subsection*{Density}

\begin{figure}
\centering
 {\includegraphics[width=0.45\textwidth] {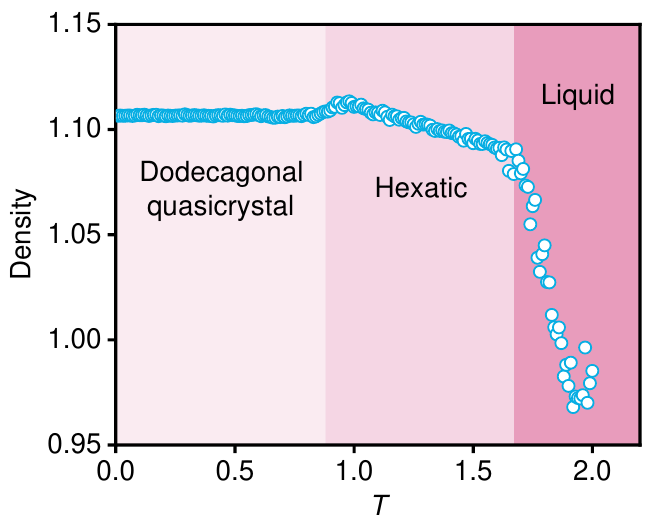}}
  \caption{Temperature dependence of the density, highlighting two distinct phase transitions as the system undergoes cooling, which link the liquid phase, hexatic phase, and quasicrystal phase.
  }
  \label{fig:12}
\end{figure}

During the rapid cooling process from the high-temperature liquid phase, the density displays two distinct thermodynamic transitions, as illustrated in Fig.~\ref{fig:12}. The first transition occurs at approximately $T_{LH} = 1.78$, marking the shift from the liquid phase to the orientationally correlated hexatic phase. Subsequently, the second transition takes place as the system undergoes a transformation from the hexatic phase to a dodecagonal quasicrystal phase, this apparent dynamical transition occurring at around $T = 0.87$. It is worth noting that the transition from a hexatic to a quasicrystal phase is characterized by a positive thermal expansion. In contrast, during the transition from the liquid phase to the hexatic phase, there is no observable expansion or compression of the material, as indicated in Fig.~\ref{fig:12}.

\subsection*{Heat Capacity}

\begin{figure}
\centering
 {\includegraphics[width=0.45\textwidth] {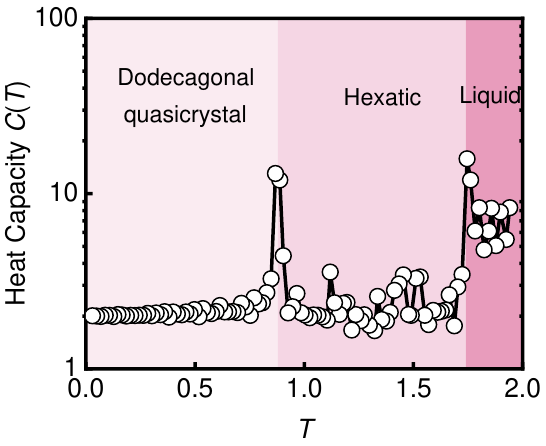}}
  \caption{Temperature dependence of heat capacity, revealing two prominent phase transitions that correlate with shifts in density, as depicted in Fig.~\ref{fig:12}.}
  \label{fig:13}
\end{figure}

We calculate the heat capacity using the ``direct" method of taking the derivative of the total energy with respect to temperature in cooling paths. Although this method, while applicable in non-equilibrium heating/cooling paths, may large uncertainties in estimating thermodynamic properties in equilibrium \cite{Bunz2014}. It is sufficient for calculating the relative value of different phases and representing phase transitions. In a manner similar to the density, the heat capacity as a function of temperature (Fig.~\ref{fig:13}) allows us to precisely estimate the location of the thermodynamic transitions of the quasicrystal material. These transition temperatures ($T_{LH} = 1.78$ and $T_{HQ} = 0.87$) are consistent with previous findings. While both of these transitions seem sharp, it is not possible to deduce the transition order due to the limited system size, as mentioned previously. 

\subsection*{Radial distribution function}

The RDF is a widely utilized tool for elucidating the structural characteristics of ordered and disordered materials. In the case of crystalline materials, $g(r)$ exhibits distinct peaks corresponding to the periodic arrangements of particles, signifying the presence of long-range order. For isotropic systems, $g(r)$ relies solely on the magnitude of the relative atomic distances and can be calculated as follows:
\begin{equation}\label{eq:A1}
g \left( r \right) = \frac{1}{2\pi \,r\,\Delta r} \frac{1}{\rho} \left \langle \sum_{i \ne i'}^{ } \delta \left( r- \lvert \mathbf{r}_{i}-\mathbf{r}_{i'} \rvert \right) \right \rangle,
\end{equation}
where $\rho$ represents the number density.

\begin{figure}
\centering
 {\includegraphics[width=0.45\textwidth] {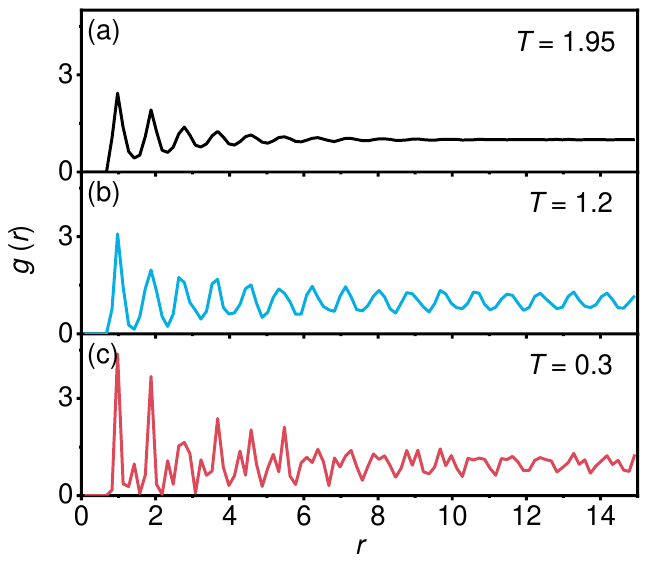}}
  \caption{Radial distribution function of (a) liquid phase at $T = 1.95$, (b) hexatic phase at $T = 1.2$ and (c) dodecagonal quasicrystal at $T = 0.3$.
  }
  \label{fig:14}
\end{figure}

In Fig.~\ref{fig:14}, we present the RDF for the liquid phase at $T = 1.95$, the hexatic phase at $T = 1.2$, and the dodecagonal quasicrystal at $T = 0.3$. The RDF of the quasicrystal exhibits a distinctive behavior that positions it between the characteristics of crystalline and disordered materials. This observation supports the notion of an intermediate state of matter, as alluded to in the title and introduction of our paper.

In addition to the well-established RDF profiles in amorphous and hexatic phases, the quasicrystal exhibits unique features. In the short range, the RDF of the quasicrystal reveals distinct peaks, which can be attributed to the quasi-periodic tiling. The first peak emerges at approximately $r = 1.0$, while the second peak appears around $r = 1.89$. These values are readily explicable as they correspond to the positions of potential energy wells (stable states) within the LJG-type potential [refer to Fig.~\ref{fig:1}(a)]. On the other hand, at longer distances the RDF of the quasicrystal displays typical characteristics of disordered materials. It gradually converges towards a constant value with non-vanishing fluctuations, reflecting the absence of long-range translational symmetry, akin to amorphous systems, while still retaining a certain translational order.

\subsection*{Static structure factor}

In MD simulations, a common method for computing $S(k)$ is through the Fourier transform of the RDF $g(r)$ \cite{HANSEN2013455}. Specifically, $S(k)$ is given by the equation:
\begin{equation}\label{eq:A2}
S(k) = 1 + 2\pi\rho \int_{0}^{\infty}\left[ g \left( r \right) -1 \right] \frac{\sin \left( kr \right) }{kr}\,r \,dr,
\end{equation}
where $\rho$ is the number density of particles, and $k$ is the wave vector. 

\begin{figure}
\centering
 {\includegraphics[width=0.45\textwidth] {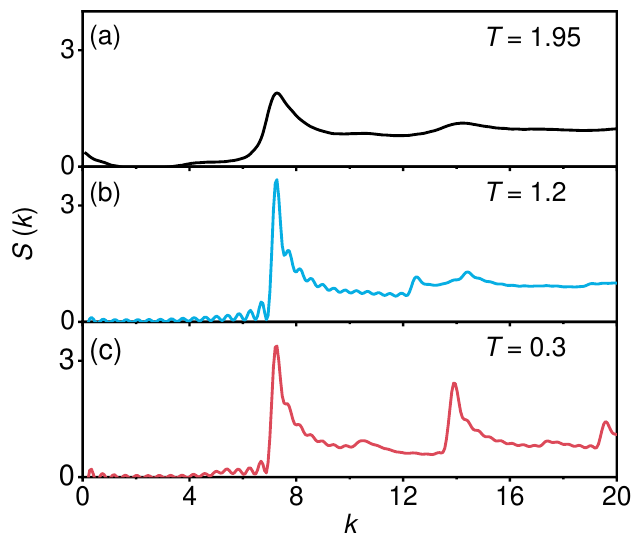}}
  \caption{Static structure factor of the (a) liquid phase at $T = 1.95$, (b) hexatic phase at $T = 1.2$ and (c) dodecagonal quasicrystal at $T=0.3$.
The rapid oscillations in (b) and (c) arise from insufficient statistics of small $k$ data with finite simulation size.
  }
  \label{fig:15}
\end{figure}

As depicted in Fig.~\ref{fig:15}, the static structure factors of the liquid, hexatic, and quasicrystal phases all prominently feature a main peak around $k^* = 2\pi/a$, where $a$ represents the nearest-neighbor distance in real space. A pronounced peak in the structure factor at $k^*$ serves as a indication of local correlations in the particle positions in the condensed material. This peak corresponds to a wave vector that generates a long-range oscillatory behavior with a periodicity of $a$ in real space \cite{Asgari2001}. Moreover, the height of the first peak in the static structure factor is a critical parameter for predicting the liquid-solid phase transition. According to the Hansen-Verlet criterion in crystalline materials the freezing transition occurs when the value of the first peak in $S(k)$ reaches a specific constant value \cite{Cheshire1969}. In the liquid phase, the first peak is notably lower than those in the hexatic and quasicrystal phases, underscoring both the lack of long-range order and the inherent disorder in the system \cite{Li2021}.

\section*{Appendix B: Collective motion in dodecagonal quasicrystal}

\begin{figure}
\centering
 {\includegraphics[width=0.45\textwidth] {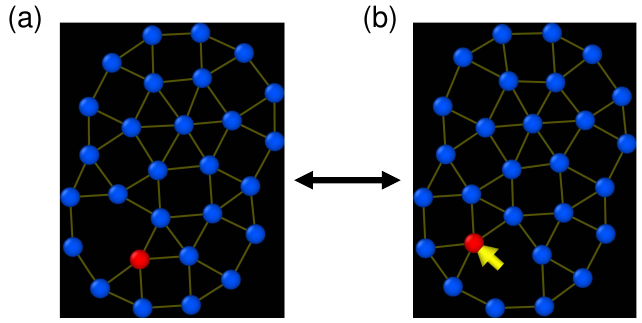}}
  \caption{Illustration of defect hopping at low temperature $T = 0.3$. (a) and (b) are the spatial distribution of a particle cluster at $t = 0$ and $t = 3 \times 10^4$, respectively. The yellow arrow represents the displacement vector of the red particle which has been scaled 1.5 times for clear observation. 
A supplementary video 1 is provided to visualize the dynamic process of such defect hopping.
  }
  \label{fig:16}
\end{figure}

We investigate the particle displacement process during fast $\beta$ relaxation at lower temperatures. In Fig.~\ref{fig:16}, we present an illustration of a putative defect flip at the very low temperature, $T = 0.3$. Notably, all observed jumps occur at structural imperfections within the quasicrystal. In the case of a hexagonal tile, nonadjacent particles exist in a relatively unstable state, as predicted by the LJG potential. Consequently, particles apparently tend to move in search of more energetically favored positions, providing a possible energetic impetus for defect flips. Fig.~\ref{fig:16}(a) and \ref{fig:16}(b) depict the spatial distribution of the same particle cluster at $t = 0$ and $t = 3 \times 10^4$, respectively. The red atom undergoes a ``jump" during this period. Interestingly, regardless of how many times these putative defect flips occur, the form of defects, \textit{i.e.}, the hexagonal tiles, persists and no topological transition is observed. The whole dynamic process is shown in Supplemental Video 2. Moreover, the number of defects exhibits a strong correlation with the preparation process of the quasicrystal phase, specifically the cooling rate. Quasicrystals with lower potential energy, corresponding to a lower concentration of defects, can be obtained through a slower cooling rate. In practice, forming ``perfect" quasicrystals without defects is challenging due to the limitations of the timescale and length scale of our atomistic simulations, making it difficult to reach an equilibrium concentration of defects.

\section*{Appendix C: Energy landscape of atomic motion}

\begin{figure*}
\centering
 {\includegraphics[width=1\textwidth] {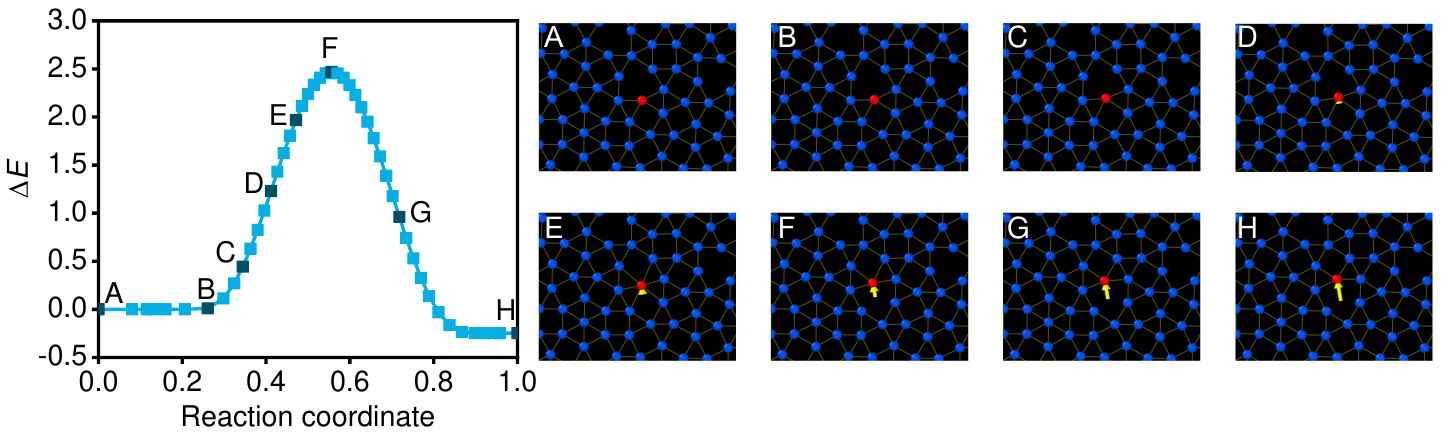}}
  \caption{Minimum energy pathway of a defect flip. The yellow arrows represent displacement vectors which are scaled twice for clear observation. The atomistic configurations from 'A' to 'H' shown in the right panels are corresponding to positions of the same letters indicated on the energy pathway in the left panel.
  }
  \label{fig:17}
\end{figure*}

\begin{figure*}
\centering
 {\includegraphics[width=1\textwidth] {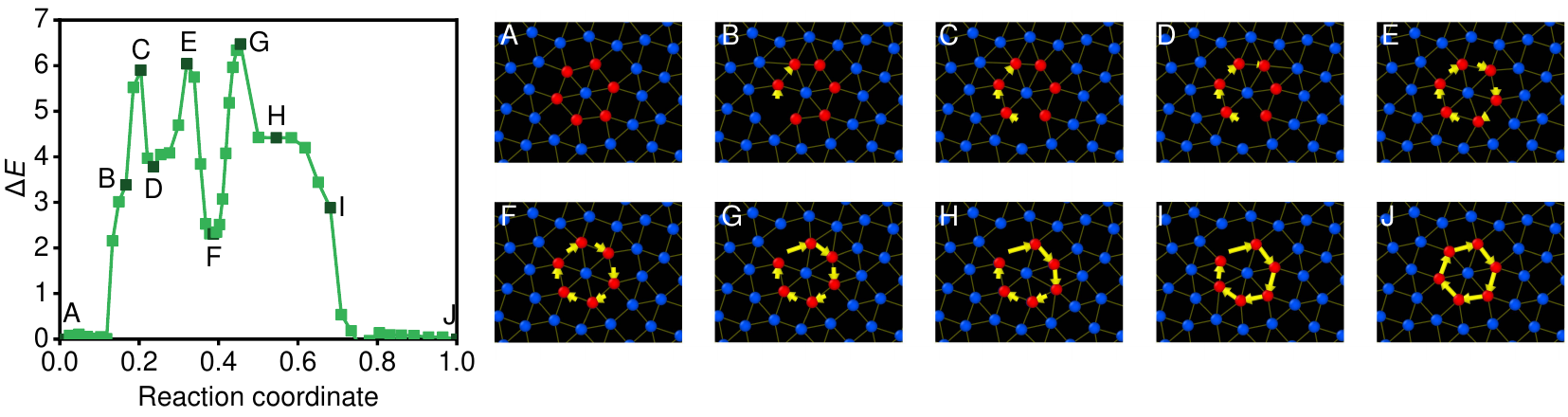}}
  \caption{Minimum energy pathway of a string-like motion in form of a single loop. The yellow arrows represent the real displacement vectors. The atomistic configurations from 'A' to 'J' shown in the right panels are corresponding to positions of the same letters indicated on the energy pathway in the left panel.
  }
  \label{fig:18}
\end{figure*}

\begin{figure*}
\centering
 {\includegraphics[width=1\textwidth] {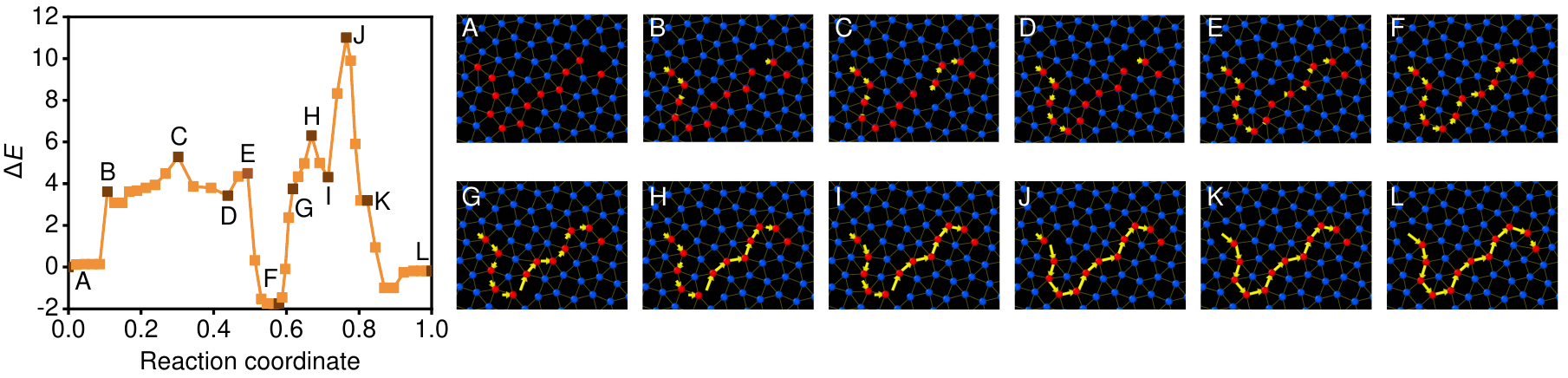}}
  \caption{Minimum energy pathway of a string-like motion in form of a crooked line. The yellow arrows represent the real displacement vectors. The atomistic configurations from 'A' to 'L' shown in the right panels are corresponding to positions of the same letters indicated on the energy pathway in the left panel.
  }
  \label{fig:19}
\end{figure*}

Given the unique atomic motion mechanisms observed in quasicrystals, we further analyze these two particle displacement modes from an energetic perspective to obtain a deeper insight into this striking phenomenon. We next compute both the activation energy and the corresponding minimum energy pathway (MEP) for both phason flip and string-like motion using the nudged elastic band (NEB) method \cite{Henkelman2000}.

First, Fig.~\ref{fig:17} illustrates the MEP of what we tentatively identify as a phason flip. The displacement process apparently involves an atom climbing over a single energy saddle point. We select specific characteristic points along the MEP to observe the detailed atomic displacement form. As the energy increases, atoms gradually move into the position of an existing defect. Upon reaching the saddle point F, a new form of defect appears. Subsequently, the atom gradually settles into a new equilibrium position. The activation energy required for this entire process is approximately 2.5, aligning well with the data obtained from the Arrhenius law in Fig.~\ref{fig:7} at the low-temperature regime.

In the case of string-like motion, depicted in Figs.~\ref{fig:18} and \ref{fig:19}, the energetic landscape traversed evidently involves a much greater complexity, the MEP involves multiple saddle points and several meta-stable positions. The illustrative ring-like exchange motion initiates with the movements of two or three particles of the hexagonal ring, accompanied by the generation of a defect left in the tail (Fig.~\ref{fig:18}D). This leads to the formation of an initial meta-stable state, with $\Delta E$ for this process being about 6, nearly three times the value assigned to the putative phason flip. Subsequently, it evolves into a semi-loop form (Fig.~\ref{fig:18}E) with the disappearance of the produced defect in Fig.~\ref{fig:18}C. This cluster tends to be also meta-stable, causing the MEP to drop to a point F. At this point, the structure and bonds differ slightly from the initial configuration after traveling half the distance of the nearest neighbor separation, with the energy of state F being higher than that in the initial state. No further meta-stable state is encountered after this stage in the collective motion evolution, apparently prompting the cluster of atoms to continue its motion in a collective way. Ultimately, a full loop is ``completed", and the structure returns to its initial equilibrium position.

For an open string, as illustrated in Fig.~\ref{fig:19}, a similar phenomenon to the ring-like string is observed. This type of string starts with the motion of only two or three particles (B and C), accompanied by the generation and annihilation of defects in arbitrary locations in the string. A near linear particle displacement path is observed at F where every particle moves half the distance of the nearest neighbor separation, culminating in the completion of a full string. At the end of the string ``lifetime", every atom has traveled the nearest-neighbor distance, and defects are left as ``debri" from this form of collective motion. Throughout the process, the saddle point appears at point J, with an energy barrier of more than 10.


\bibliography{reference.bib}

\end{document}